\documentclass[11pt]{article}

\usepackage{authblk}  % have affiliations with article
\usepackage{graphicx} % Include figure files
\usepackage{epsfig}
\usepackage{dcolumn}  % Align table columns on decimal point
\usepackage{bm}       % bold math
\usepackage{amsmath}
\usepackage{amssymb}

\newcommand{\beq}{\begin{equation}} \newcommand{\eeq}{\end{equation}}
\newcommand{\bea}{\begin{eqnarray}} \newcommand{\eea}{\end{eqnarray}}

  \newcommand
{\Romannumeral}[1]{\uppercase\expandafter{\romannumeral#1}}

\newcommand{\be}{\begin{enumerate}} \newcommand{\ee}{\end{enumerate}}
\newcommand{\bi}{\begin{itemize}} \newcommand{\ei}{\end{itemize}}
\newcommand{\ba}{\begin{array}} \newcommand{\ea}{\end{array}}
\newcommand{\bc}{\begin{center}} \newcommand{\ec}{\end{center}}
\newcommand{\bt}{\begin{tabular}} \newcommand{\et}{\end{tabular}}

\def\lsim{\mathrel{\rlap{\lower4pt\hbox{\hskip1pt$\sim$}}
    \raise1pt\hbox{$<$}}}           % less than or approx. symbol
\def\gsim{\mathrel{\rlap{\lower4pt\hbox{\hskip1pt$\sim$}}
    \raise1pt\hbox{$>$}}}           % greater than or approx. symbol

\newcommand{\half}{\textstyle {1\over2} \displaystyle}    % One half
\newcommand{\third}{\textstyle {1\over3} \displaystyle}   % One third
\newcommand{\quarter}{\textstyle {1\over4} \displaystyle} % One quarter
\newcommand{\Dslash}{{\hbox{D}\kern-0.6em\raise0.15ex\hbox{/}}} % D slash

\renewcommand{\et}{\eta}

\begin{document}

\setlength{\oddsidemargin}{0cm} \setlength{\baselineskip}{7mm}

\begin{normalsize}\begin{flushright}

% CERN-PH-TH/2006-147 \\
% DAMTP-2006-59 \\
% UCI-2004-xx \\
December 2012 \\

\end{flushright}\end{normalsize}

\begin{center}
  
\vspace{5pt}

{\Large \bf Wheeler-DeWitt Equation in 3 + 1 Dimensions }

\vspace{30pt}

{\sl Herbert W. Hamber}
$^{}$\footnote{e-mail address : Herbert.Hamber@aei.mpg.de} 
\\
Max Planck Institute for Gravitational Physics \\
(Albert Einstein Institute) \\
D-14476 Potsdam, Germany\\

\vspace{10pt}

{\sl Reiko Toriumi}
$^{}$\footnote{e-mail address : RToriumi@uci.edu} \\
Department of Physics and Astronomy, \\
University of California, \\
Irvine, California 92697-4575, USA \\

\vspace{5pt}

and

\vspace{5pt}

{\sl Ruth M. Williams}
$^{}$\footnote{e-mail address : R.M.Williams@damtp.cam.ac.uk} \\
Department of Applied Mathematics and Theoretical Physics, \\
University of Cambridge,\\
Wilberforce Road, Cambridge CB3 0WA, United Kingdom, \\

and 

Girton College, University of Cambridge, Cambridge CB3 0JG, United Kingdom. \\

\vspace{10pt}

\end{center}

\begin{center} {\bf ABSTRACT } \end{center}

\noindent

Physical properties of the quantum gravitational vacuum state
are explored by solving a lattice version of the 
Wheeler-DeWitt equation.
The constraint of diffeomorphism invariance is strong enough to uniquely
determine part of the structure of the vacuum wave functional in the limit of 
infinitely fine triangulations of the three-sphere.
In the large fluctuation regime the nature of the wave function solution
is such that a physically acceptable ground state emerges,
with a finite non-perturbative correlation length naturally 
cutting off any infrared divergences.
The location of the critical point in Newton's constant $G_c$,
separating the weak from the strong coupling phase,
is obtained, and it is inferred from the general structure of the wave functional
that fluctuations in the curvatures become unbounded at this point.
Investigations of the vacuum wave functional further suggest 
that for weak enough 
coupling, $G< G_c$, a pathological ground state with no continuum limit appears,
where configurations with small curvature have vanishingly small probability.
One would then be lead to the conclusion that the weak coupling, perturbative ground
state of quantum gravity is non-perturbatively unstable, and that
gravitational screening cannot be physically realized in the lattice theory.
The results we find tend to be in general agreement with the Euclidean
lattice gravity results, and would suggest that the Lorentzian 
and Euclidean lattice formulations of gravity ultimately
describe the same underlying non-perturbative physics.

% \vspace{15pt}

% \begin{center} {\it (Submitted to the Physical Review D)} \end{center}
% \noindent PACS Numbers: 04.60.-m, 04.60.Gw, 04.60.Nc, 98.80.Qc
% Above are: Quant Grav, Cov Sum over Hist Quant, Latt Grav, Quant Cosm 

\vfill

%% no page number on 1-st page
\pagestyle{empty}

\newpage

\pagestyle{plain}

\section{Introduction}
\label{sec:intro}

\vskip 20pt

We have argued in previous work that the correct identification 
of the true ground
state for quantum gravitation necessarily requires the introduction of a
consistent nonperturbative cutoff, followed by the construction of 
the continuum limit in accordance with the methods
of the renormalization group.
To this day the only known way to introduce such a non-perturbative
cutoff reliably in quantum field theory is via the lattice formulation.
A wealth of results have been obtained over the years using
the Euclidean lattice formulation, which allows the identification
of the physical ground state and the accurate calculations of
gravitational scaling dimensions, relevant for the scale 
dependence of Newton's constant in the universal scaling limit.
 
In this work we will focus instead on the Hamiltonian approach to gravity,
which assumes from the very beginning a metric with Lorentzian signature.
Recently a Hamiltonian lattice formulation was written down
based on the Wheeler-DeWitt equation, where the gravity
Hamiltonian is expressed in the metric-space representation.
Specifically, in \cite{hw11,htw12} a general discrete Wheeler-DeWitt 
equation was given for pure gravity, based on the simplicial lattice 
transcription of gravity formulated by Regge and Wheeler.
Here we extend the work initiated in \cite{hw11,htw12} to the
physical case of $3+1$ dimensions,
and show how nonperturbative vacuum solutions to the lattice Wheeler-DeWitt
equations can be obtained for arbitrary values of Newton's constant $G$.
The procedure we follow is similar to what was done earlier in $2+1$
dimensions.
We solve the lattice equations exactly for
several finite and regular triangulations of the three-sphere, and then 
extend the result to an arbitrarily large number of tetrahedra.
We then argue that for large enough volumes the exact lattice wave functional is expected
to depend on geometric quantities only, such as the total volumes and the total
integrated curvature.
In this process, the regularity condition on the solutions of the wave equation at
small volumes plays an essential role in constraining 
the form of the vacuum wave functional.
A key ingredient in the derivation of the results is of course the local 
diffeomorphism invariance of the Regge-Wheeler lattice formulation.

From the structure of the resulting wave function a 
number of potentially useful physical results can be obtained.
First one observes that the non-perturbative correlation 
length is found to be finite for sufficiently large $G$.
At the critical point $G=G_c$, which we determine exactly
from the structure of the wave function,
fluctuations in the curvature become unbounded, thus signaling
a divergence in the non-perturbative gravitational correlation length.
We argue that such a result can be viewed as consistent with the existence 
of a non-trivial ultraviolet fixed point 
(or a phase transition in statistical field theory language) in $G$.
Furthermore, the behavior of the theory in the vicinity of such a
fixed point is expected to determine, through standard 
renormalization group arguments, the scale dependence of the 
gravitational coupling in the vicinity of the ultraviolet fixed point.

An outline of the paper is as follows.
In Sec. 2, as a background to the rest of the paper, 
we briefly summarize the formalism of canonical gravity.
At this stage the continuum Wheeler-DeWitt equation with its invariance
properties are introduced.
We then briefly outline the general properties of the lattice 
Wheeler-DeWitt equation presented in our previous work,
and in Sec. 3 we make explicit various quantities that appear
in it.
Here we also emphasizes the important role of 
continuous lattice diffeomorphism invariance in the Regge theory,
as it applies to the case of $3+1$-dimensional gravity.
Sec. 4 focuses on basic scaling properties of the lattice equations
and useful choices for the lattice coupling constants,
with the aim of giving a more transparent form to the 
results obtained later.
Sec. 5 presents an outline of the method 
of solution for the lattice equations,
which are later discussed in some detail for a number of 
regular triangulations of the three-sphere.
Then a general form of the wave function is given that covers all
previous discrete cases, and thus allows a study of the 
infinite volume limit.
Sec. 6 discusses the issue of how to define an average
volume and thus an average lattice spacing, an
essential ingredient in the interpretation of the results given later.
Sec. 7 discusses modifications of the wave function solution
obtained when the explicit curvature term in the Wheeler-DeWitt
equation is added.
Later a partial differential equation for the wave function
is derived in the curvature and volume variables.
General properties of the solution to this equation are
discussed in Sec. 8.
Sec. 9 contains a brief summary of the results obtained so far.

\vskip 40pt

\section{Continuum and Discrete Wheeler-DeWitt Equation}

\label{sec:wdweq}

\vskip 20pt

Our work deals with the canonical quantization of gravity, and
we begin here therefore with a very brief summary of the classical 
canonical formalism \cite{dir58} as formulated by Arnowitt, Deser and Misner 
\cite{adm62}.
Many of the results found in this section are not new, but
nevertheless it will be useful, in view of later applications, 
to recall here the main results and provide suitable references for 
expressions used in the following sections.
Here $x^i$ $(i=1,2,3)$ will be coordinates on a three-dimensional
manifold, and indices will be raised and lowered with 
$g_{ij} ( {\bf x} ) $ $(i,j=1,2,3)$, the three-metric 
on the given spacelike hypersurface.
As usual $g^{ij}$ denotes the inverse of the matrix $g_{ij}$.
Our conventions are such that the space-time metric has 
signature $-+++$, that ${}^4 \! R $ is non-negative in a 
space-time containing normal matter, and that 
$ {}^3 \! R $ is positive in a 3-space of positive curvature.

One goes from the classical to the quantum description 
of gravity by promoting the metric
$g_{ij}$, the conjugate momenta $\pi^{ij}$, 
the Hamiltonian density $H$ and the momentum density $H_i$ 
to quantum operators, with ${\hat g}_{ij}$ and ${\hat \pi}^{ij}$ 
satisfying canonical commutation relations.
Then the classical constraints select physical states
$ \vert \Psi \rangle $, such that in the absence of sources
\beq
{\hat H} \, \vert \Psi \rangle 
\, = \, 0
\;\;\;\;\;\;
{\hat H}_i \, \vert \Psi \rangle 
\, = \, 0 \;\;  ,
\eeq
whereas in the presence of sources one has more generally
\beq
{\hat T} \, \vert \Psi \rangle 
\, = \, 0
\;\;\;\;\;\;
{\hat T}_i \, \vert \Psi \rangle 
\, = \, 0 \;\; ,
\label{eq:quant_const}
\eeq
with ${\hat T}$ and ${\hat T}_i$ describing matter
contributions that can be added to ${\hat H}$ and ${\hat H}_i$.
As is the case in nonrelativistic quantum mechanics, one can choose 
different representations for the canonically conjugate operators 
${\hat g}_{ij}$ and ${\hat \pi}^{ij}$.
In the functional metric representation one sets
\beq
{\hat g}_{ij} ( {\bf x} ) 
\; \rightarrow \;   
g_{ij} ( {\bf x} ) 
\;\;\;\;\;\;\;
{\hat \pi}^{ij} ( {\bf x} ) 
\; \rightarrow \;  
- i \hbar \cdot16 \pi G \cdot
{ \delta \over \delta g_{ij} ( {\bf x} )  } \; .
\eeq
Then quantum states become wave functionals of the three-metric
$ g_{ij} ({\bf x} ) $,
\beq
\vert \Psi \rangle \; \rightarrow \;  \Psi \, [ g_{ij} ( {\bf x} ) ] \; .
\eeq
The constraint equations in Eq.~(\ref{eq:quant_const})
then become the Wheeler-DeWitt equation \cite{whe64,dew64}
\beq
\left \{ -  \, 16 \pi G \cdot G_{ij,kl}  \, 
{ \delta^2 \over \delta g_{ij} \, \delta g_{kl}  } \, - \,
{ 1 \over 16 \pi G } \, \sqrt{g} \,
\left ( \,  {}^3 \! R \, - \, 2 \lambda \, \right )
\, + \,  {\hat H}^\phi \right \} \; \Psi [ g_{ij} ( {\bf x} ) ] \, = \, 0 \; ,
\label{eq:wd_1}
\eeq
and the momentum constraint equation listed below.
In Eq.~(\ref{eq:wd_1}) $ G_{ij,kl} $ is the 
inverse of the DeWitt supermetric,
\beq
G_{ij,kl} \, = \,  
\half \, g^{-1/2} \left ( 
g_{ik} g_{jl} + g_{il} g_{jk} \, 
- \, g_{ij} g_{kl} \right ) \;\; .
\label{eq:dew_inv}
\eeq
The three-dimensional DeWitt supermetric itself is given by
\beq
G^{ij,kl} \, = \,  
\half \, \sqrt{g} \, 
\left ( g^{ik} g^{jl} + g^{il} g^{jk} 
\, - \, 2 \, \, g^{ij} g^{kl} \right ) \;\; .
\label{eq:dew}
\eeq
In the metric representation the diffeomorphism constraint reads
\beq
\left \{ 2 \, i  \, g_{ij}  \, 
\nabla_k \, { \delta \over \delta g_{jk}  } 
\, + \,
{\hat H}^\phi_i \right \} \; 
\Psi [ g_{ij} ( {\bf x} )  ] \, = \, 0 \; ,
\label{eq:wd_2}
\eeq
where $ {\hat H}^\phi$ and $ {\hat H}^\phi_i $ again are possible matter
contributions.
In the following, we shall set both of these to zero as we will focus here
almost exclusively on the pure gravitational case.
Then the last contraint represents the necessary and sufficient
condition that the wave functional $\psi [g]$ be an invariant
under coordinate transformations \cite{dew67}.

% NEW paragraph :

We note here that in the continuum one expects the commutator of 
two Hamiltonian constraints to be related to the diffeomorphism
constraint.  
In the following we will, for the time being, overlook this rather
delicate issue, and focus our efforts instead mainly on the solution of the
explicit (lattice) Hamiltonian constraint of Eq.~(\ref{eq:wd_latt1}).  
It should nevertheless be possible to revisit this important issue
at a later stage, once an exact, or approximate, candidate expression 
for the wave functional is found. The key issue at that stage will
then be if the lattice wave functional satisfies all physical requirements, 
including the momentum constraint, in a suitable lattice scaling limit 
wherein the (average) lattice spacing is much smaller than a suitable 
physical scale, such as the scale of the local curvature, or some
other sort of agreeable physical correlation length. For a more in-depth 
discussion of the analogous problem in $2+1$ dimensions we refer the 
reader to our previous work \cite{htw12}, where an explicit form for 
the candidate wave functional was eventually given in terms of 
manifestly invariant quantities such as areas and curvatures.

We should also mention here that a number of rather basic issues need 
to be considered before one can 
gain a consistent understanding of the full content of the theory 
[see, for example, \cite{leu64,kuc76,kuc92,ish93,bar98}].
These include potential problems with operator ordering, 
and the specification of a suitable Hilbert space, 
which entails a choice for the norm
of wave functionals,  for example in the Schr\"odinger form
\beq
\lVert \Psi \rVert^2 \, = \, 
\int  d \mu [g] \; \Psi^{*} [ g_{ij} ] \; \Psi [ g_{ij} ] \; ,
\label{eq:norm}
\eeq
where $ d \mu [g] $ is the appropriate (DeWitt) 
functional measure over the three-metric $g_{ij}$.
In this work we will attempt to address some of those issues, 
as they will come up within the relevant calculations.

In this paper the starting point will be the Wheeler-DeWitt equation 
for pure gravity in the absence of matter, Eq.~(\ref{eq:wd_1}), 
\beq
\left \{ \, -  \, (16 \pi G )^2 \, 
G_{ij,kl}  ( {\bf x} )  \, 
{ \delta^2 \over \delta g_{ij} ( {\bf x} ) \, 
\delta g_{kl}  ( {\bf x} ) } \, - \,
\sqrt{g ( {\bf x} ) } \; \left ( \;  
{}^3 \! R ( {\bf x} ) \, - \, 2 \lambda \, \right ) \, \right \} \; 
\Psi [ g_{ij} ( {\bf x} ) ] 
\, = \, 0 \; ,
\label{eq:wd_1a}
\eeq
combined with the diffeomorphism constraint of Eq.~(\ref{eq:wd_2}),
\beq
\left \{ \, 2 \, i  \, 
g_{ij}  ( {\bf x} ) \, \nabla_k ( {\bf x} ) \,
 { \delta \over \delta g_{jk}  ( {\bf x} ) }  \, 
\right \} \; 
\Psi [ g_{ij} ( {\bf x} )  ] 
\, = \, 0 \; .
\label{eq:wd_2a}
\eeq
Both of these equations express a constraint on the state 
$ \vert \Psi \rangle$ at {\it every} ${\bf x}$.
It is then natural to view Eq.~(\ref{eq:wd_1a})
as made up of
three terms, the first one identified as the kinetic term
for the metric degrees of freedom,
the second one involving $ - \, \sqrt{g} \, {}^3 \! R $ and thus
seen as a potential energy contribution
(of either sign, due to the nature of the 3-curvature $ {}^3 \! R $), 
and finally the cosmological constant term proportional
to $ + \, \lambda\, \sqrt{g} $ acting as a mass-like term.
The kinetic term contains a Laplace-Beltrami-type
operator acting on the 6-dimensional Riemannian manifold
of positive definite metrics $g_{ij}$, with 
$G_{ij,kl}$ acting as its contravariant metric \cite{dew67}.
As shown in the quoted reference, the manifold
in question has hyperbolic signature $-+++++$, with pure
dilations of $g_{ij}$ corresponding to timelike displacements
within this manifold of metrics.

Next we turn to the lattice theory.
Here we will generally follow the procedure outlined in \cite{hw11}
and discretize the continuum Wheeler-DeWitt 
equation directly, a procedure that makes sense in the lattice formulation, 
as these equations are still formulated in terms of geometric objects, 
for which the Regge theory is very well suited.
It is known that on a simplicial lattice \cite{rowi81,che82,lee83,itz83,hw84,lesh84,har85} 
(see for example \cite{hbook} for a more detailed presentation of the 
Regge-Wheeler lattice formulation)
deformations of the squared edge lengths are linearly 
related to deformations of the induced metric.
In a given simplex $\sigma$, take coordinates based at a vertex $0$, 
with axes along the edges emanating from $0$.
Then the other vertices are each at unit coordinate distance from $0$ 
(see Figure 1 as an example of this labeling for a tetrahedron). 
With this choice of coordinates, the metric within a given simplex is
\beq
g_{ij} (\sigma) \, = \, 
\half \, 
\left ( l_{0i}^2 + l_{0j}^2 - l_{ij}^2 
\right ) \;\; .
\label{eq:latmet}
\eeq
We note that in the following discussion only edges and 
volumes along the spatial directions are involved.
Then by varying the squared volume of a given simplex 
$\sigma$ in $d$ dimensions
to quadratic order in the metric (in the continuum), or in the squared 
edge lengths belonging to that simplex (on the lattice), one is led to
the identification \cite{lun74,har97}
\beq
G^{ij} (l^2) \, = \, 
- \; d! \; \sum_{\sigma} \;
{1 \over V (\sigma)} \; 
{ \partial^2 \; V^2 (\sigma) 
\over \partial l^2_i \; \partial l^2_j } \;\; ,
\label{eq:lund_regge}
\eeq
where the quantity $G^{ij} (l^2) $ is local, since the sum over $\sigma$ 
only extends over those simplices which contain either the $i$ or the $j$ edge.
In the formulation of \cite{hw11} it will be adequate
to limit the sum in Eq.~(\ref{eq:lund_regge}) to a single
tetrahedron, and define the quantity $G^{ij}$ for that
tetrahedron.
Then, in schematic terms, the lattice Wheeler-DeWitt equation for pure
gravity takes on the form
\beq
\left \{ \, -  \, (16 \pi G)^2 \, 
G_{ij} ( l^2 ) \, 
{ \partial^2 \over \partial l^2_{i} \, \partial l^2_{j}  }
\, - \,
\sqrt{g (l^2) } \; 
\left [ \; 
{}^3 \! R (l^2) \, - \, 2 \lambda \; 
\right ] \; \right \} \,
\Psi [ \, l^2 \, ] \, = \, 0 \;\; ,
\label{eq:wd_latt}
\eeq
with $G_{ij} (l^2)$ the inverse of the matrix $G^{ij} (l^2)$ given above.
The range of summation over the indices $i$ and $j$ 
and the appropriate expression for the scalar curvature
will be made explicit later in Eq.~(\ref{eq:wd_latt1}).

It is clear that Eqs.~(\ref{eq:wd_1}) or (\ref{eq:wd_latt})
express a constraint at each \lq\lq point" in space.
Indeed, the first term in Eq.~(\ref{eq:wd_latt})
contains derivatives with respect to edges $i$ and $j$ 
connected by a matrix element $ G_{ij} $ which is nonzero only if $i$ and $j$
are close to each other and thus nearest neighbor.
% NEW footnote added :
\footnote{
In Regge gravity space time diffeomorphisms correspond to movements 
of the vertices which leave the 
local geometry unchanged (see for example \cite{rowi81,har85,gauge}, 
and further references therein). In the present case the lattice 
Hamiltonian constraints can be naturally viewed as generating local 
deformations of the spatial lattice hypersurface. 
One would therefore expect the Hamiltonian constraint to be based 
here on the lattice vertices as well. 
But this seems nearly impossible to implement, as the definition 
of the local lattice supermetric $G^{ij} (l^2) $ based on 
Eq.~(\ref{eq:latmet}) clearly requires the consideration of a 
full tetrahedron, as  do the derivatives with respect to the edges, 
and finally the very definition of the curvature and volume terms. 
One could possibly still insist on defining the Hamiltonian constraint 
on a vertex by averaging over contributions from many neighboring 
tetrahedra, but this would make the lattice problem 
intractable from a practical point of view.
How this choice will ultimately affect the counting of degrees of 
freedom is unclear at this stage, for two reasons. 
The first one is that in the Regge theory there is in general a 
certain redundancy of degrees of freedom \cite{rowi81}, with 
unwanted ones either decoupling or acquiring a mass 
of the order of the ultraviolet cutoff. Furthermore, as will be shown 
later for example in Eq.~(\ref{eq:n0-n3}), the detailed relationship 
between the number of lattice vertices and tetrahedra clearly 
depends on  the chosen lattice structure, and more specifically on 
the local lattice coordination number.}
One expects therefore that the first term can be represented
by a sum of edge contributions, all from within one
($d-1$)-simplex $\sigma$ (a tetrahedron in three dimensions).
The second  term containing ${}^3 \! R (l^2)$ in Eq.~(\ref{eq:wd_latt}) 
is also local in the edge lengths:
it only involves those edge lengths which enter 
into the definition of areas, volumes and angles around the point 
${\bf x}$.
The latter is therefore described, through the deficit angle $\delta_h$, 
by a sum over contributions over all ($d-3$)-dimensional hinges 
(edges in 3+1 dimensions) $h$ attached to the simplex $\sigma$.
This then leads in three dimensions to a more explicit form
of Eq.~(\ref{eq:wd_latt})
\beq
\left \{ \, -  \, (16 \pi G )^2 
\sum_{ i,j \subset \sigma}
\, G_{ij} \, ( \sigma ) \, 
{ \partial^2 \over \partial l^2_{i} \, \partial l^2_{j}  }
\, - \, 2 \, n_{\sigma h} \; 
\sum_{ h \subset \sigma} \, l_h \, \delta_h 
\, + \, 2 \lambda  \; V_\sigma  \,
\right \} \,
\Psi [ \, l^2 \, ] \, = \, 0 \;\; .
\label{eq:wd_latt1}
\eeq
In the above expression $\delta_h$ is the deficit angle at the hinge (edge)
$h$, $l_h$ the corresponding edge length, 
and $V_\sigma = \sqrt{ g(\sigma)} $ the volume of the simplex 
(tetrahedron in three spatial dimensions) labeled by $\sigma$.
The matrix $ G_{ij} \, ( \sigma ) $ is obtained either from 
Eq.~(\ref{eq:lund_regge})
or from the lattice transcription of Eq.~(\ref{eq:dew_inv})
\beq
G_{ij,kl} \, (\sigma) \, = \,  
\half \, g^{-1/2} (\sigma) \, \left [ 
\, g_{ik} (\sigma) \, g_{jl} (\sigma) 
+ g_{il} (\sigma) \, g_{jk} (\sigma) - 
\, g_{ij} (\sigma) \, g_{kl} (\sigma) \, \right ] \; ,
\label{eq:dewitt_inv_1}
\eeq
with the induced metric $g_{ij} \, (\sigma) $ within a simplex $\sigma$ given in 
Eq.~(\ref{eq:latmet}).
Note that the combinatorial factor $ n_{\sigma h} $ gives the correct
normalization for the curvature term, since the latter 
has to give the lattice version of
$ \int \sqrt{g} \; {}^3 \! R = 2 \sum_h \delta_h l_h $  
when summed over all simplices $\sigma$.
One can see then that the inverse of $ n_{\sigma h} $ counts
the number of times the same hinge appears in 
various neighboring simplices, and depends therefore on the 
specific choice of underlying lattice structure.
The lattice Wheeler-DeWitt equation given in Eq.~(\ref{eq:wd_latt1})
was the main result of a previous paper \cite{hw11}
and was studied extensively in $2+1$ dimensions in 
previous work \cite{htw12}.

\vskip 40pt

\section{Explicit Setup for the Lattice Wheeler-DeWitt Equation}

\label{sec:setup}

\vskip 20pt

In the following we will now focus on a three-dimensional lattice made up
of a large number of tetrahedra, with squared edge lengths 
considered as the fundamental degrees of freedom.
For ease of notation, we define $l_{01}^2=a, \, l_{12}^2=b, \, l_{02}^2=c, \,
l_{03}^2=d, \, l_{13}^2=e, \, l_{23}^2=f$.
For the tetrahedron labeled as in Figure 1, we have
\beq 
g_{11} \, = \, a \, , \;\;\;\; g_{22} \, = \, c \, , \;\;\; g_{33} \,
= \, d \, ,
\eeq
\beq
g_{12} \, = \, \frac {1} {2} (a \, + \, c \, - \, b) \, , \;\;\;
g_{13} \, = \, \frac {1} {2} (a \, + \, d \, - \, e) \, , \;\;\;
g_{23} \, = \, \frac {1} {2} (c \, + \, d \, - \, f) \; ,
\eeq
and its volume $V$ is given by 
\bea
V^2 & = & \; \frac {1} {144} [ \, af (-a-f+b+c+d+e) 
\, + \, bd (-b-d+a+c+e+f) \, + \, 
\nonumber \\
&\;\;\;\; + & ce (-c-e+a+b+d+f) \, - \, abc \, - \, ade \, - \, bef \, - \, cdf ] \; .
\eea
The matrix $G^{ij}$ is then given by
\beq
G^{ij} \, = \, - \, \frac {1} {24 V}
\left (  
\begin{matrix}
-2f & e+f-b & b+f-e & d+f-c & c+f-d & p \cr
e+f-b & -2e & b+e-f & d+e-a & q & a+e-d \cr
b+f-e & b+e-f & -2b & r & b+c-a & a+b-c \cr
d+f-c & d+e-a & r & -2d & c+d-f & a+d-e \cr
c+f-d & q & b+c-a & c+d-f & -2c & a+c-b \cr
p & a+e-d & a+b-c & a+d-e & a+c-b & -2a 
\end{matrix}
\right ) \; ,
\label{eq:Gij}
\eeq
where the three quantities $p$, $q$ and $r$ are defined as
\beq
p = -2a-2f+b+c+d+e, \;\;\; q = -2c-2e+a+b+d+f, 
\;\;\; r = -2b-2d+a+c+e+f \; .
\eeq
To obtain $G_{ij}$ one can then either invert the above expression,
or evaluate
\beq
G_{ij,kl} \, = \, \frac {1} {2 \sqrt{g}} (g_{ik} \, g_{jl} \, + \,
g_{il} \, g_{jk} \, - \, g_{ij} \, g_{kl}),
\eeq
and later replace derivatives with respect to the metric
by derivatives with respect to the squared edge lengths, as in
$ { \partial \over \partial \, g_{11} } = 
{ \partial \over \partial \, a  } 
\, + \, { \partial \over \partial \, b  }
\, + \, { \partial \over \partial \, e  } $ etc.
One finds \cite{hw11} that the matrix representing the coefficients 
of the derivatives with
respect to the squared edge lengths is the same
as the inverse of $G^{ij}$, a result that provides
a nontrivial confirmation of the correctness of
the Lund-Regge result of Eq.~(\ref{eq:lund_regge}).
Then in $3+1$ dimensions, the discrete Wheeler-DeWitt equation is 
\beq
\left \{ \, - \, (16 \pi G)^2 \, G_{ij} \, {\partial^2 \over
\partial s_i \partial s_j} \, - \, 2 \, n_{\sigma h} \, \sum_{h}
\sqrt{s_h} \, \delta_h \, + \, 2 \lambda \, V \right \} \Psi [ \, s \, ] 
\, = \, 0 \; ,
\label{eq:wd_3d}
\eeq
where the sum is over hinges (edges) $h$ in the tetrahedron, and $V$
the volume of the given tetrahedron.
Note that the above represents one equation for {\it every}
tetrahedron on the lattice.
Thus if the lattice contains $N_3$ tetrahedra, there will be $N_3$
coupled equations that will need to be solved in order to determine
$\Psi [s]$.
Note also the mild nonlocality of the equation in that the curvature term,
through the deficit angles, involves edge lengths from neighboring tetrahedra.
Of course, in the continuum the derivatives also give some very mild nonlocality.
Figure 2 gives a pictorial representation of lattices that can be used for 
numerical studies of quantum gravity in 3+1 dimensions.

\begin{figure}
\begin{center}
\includegraphics[width=0.5\textwidth]{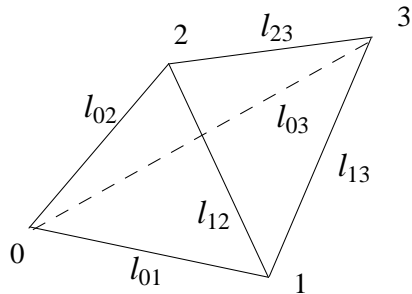}
\end{center}
\caption{
\label{fig:tet}
A tetrahedron with labels.
}
\end{figure}

\begin{figure}
\begin{center}
\includegraphics[width=0.5\textwidth]{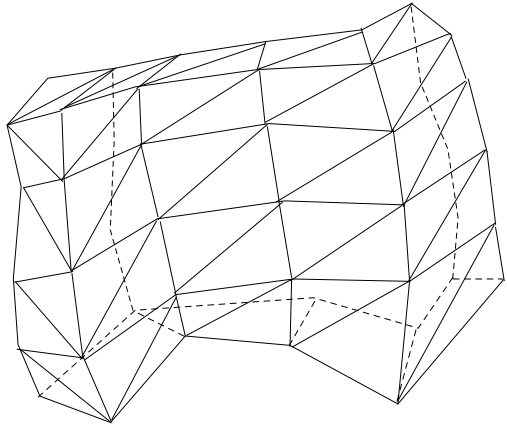}
\end{center}
\caption{
\label{fig:lattice}
A small section of a suitable spatial 
lattice for quantum gravity in $3+1$ dimensions.
}
\end{figure}

In the following we will be concerned at some point with 
various discrete, but generally regular, triangulations
of the three-sphere \cite{cox48,cox74}.
These were already studied in some detail within the framework
of the Regge theory in \cite{hw84}, where in particular
the 5-cell $\alpha_4$, the 16-cell $\beta_4$ and the 600-cell 
regular polytopes (as well as a few others) were considered
in some detail.
For a very early application of these regular triangulations
to general relativity see \cite{lin57}.

We shall not dwell here on a well-known key aspect of the 
Regge-Wheeler theory, which is the presence of a continuous, local
lattice diffeomorphism invariance, whose main aspects in regards
to its relevance for the $3+1$ formulation of gravity were
already addressed in some detail in various works, both in the
framework of the lattice weak field expansion \cite{rowi81,hw11}, 
and beyond it \cite{har85,gauge}.
Here we will limit ourselves to some brief remarks on
how this local invariance
manifests itself in the $3+1$ formulation, and, in particular, 
in the case of the discrete
triangulations of the sphere studied later on in this paper.
In general, lattice diffeomorphisms in the Regge-Wheeler theory
correspond to local deformations
of the edge lengths about a vertex, which leave the local geometry
physically unchanged, the latter being described by the values of local
lattice operators corresponding to local volumes and curvatures
\cite{rowi81,har85,gauge}.
The case of flat space (curvature locally equal to zero) or near-flat space
(curvature locally small) is obviously the simplest to analyze
\cite{gauge}: 
by moving the location of the vertices around on a smooth manifold one can
find different assignments of edge lengths representing locally the
same flat, or near-flat, geometry.
It is then easy to show that one obtains a $ d \cdot N_0$-parameter 
family of local transformations for the edge lengths, as expected
for lattice diffeomorphisms.
For the present case, the relevant lattice diffeomorphisms are
the ones that apply to the three-dimensional, spatial theory.
The reader is referred to \cite{hw93} and, more recently,
\cite{hw11} for their explicit form within the framework 
of the lattice weak field expansion.

With these observations in mind, we can now turn to a discussion
of the solution method for the lattice Wheeler-DeWitt equation
in $3+1$ dimensions.
One item that needs to be brought up at this point is the proper
normalization of various terms (kinetic, cosmological and curvature)
appearing in the lattice equation of Eqs.~(\ref{eq:wd_latt1})
and (\ref{eq:wd_3d}).
For the lattice gravity action in $d$ dimensions one has generally
the following correspondence
\beq
\int d^d x \, \sqrt{g}  \;\;\; \longleftrightarrow  
\;\;\;  \sum_{\sigma} \, V_{\sigma} \; ,
\label{eq:reg_v}
\eeq
where $V_\sigma$ is the volume of a simplex; in three dimensions it is
simply the volume of a tetrahedron.
The curvature term involves deficit angles in the discrete case,
\beq
\half \,  \int  \, d^d x \, \sqrt{g} \, R  \;\;\;  \longleftrightarrow
 \;\;\;   \sum_{h} \, V_h \, \delta_h  \; ,
\label{eq:reg_r}
\eeq
where $ \delta_h $ is the deficit angle at the hinge $h$, and $V_h$ the
associated ``volume of a hinge'' \cite{reg61}.
In four dimensions the latter is the area of a triangle (usually
denoted by $V_h$), whereas in three dimensions it is simply given by the
length $l_h$ of the edge labeled by $h$. 
In this work we will focus almost exclusively on the case of $3+1$ dimensions;
consequently the relevant formulas will be 
Eqs.~(\ref{eq:reg_v}) and (\ref{eq:reg_r}) for dimension $d=3$.

The continuum Wheeler-DeWitt equation is local, as can
be seen from Eq.~(\ref{eq:wd_1a}).
One can integrate the Wheeler-DeWitt operator over all space and obtain
\beq
\left \{ - \, \left( 16 \pi \, G \right)^2 \, 
\int d^3 x \; {\bf \Delta} (g) \, + 2 \lambda \int d^3 x \, \sqrt{g} 
- \int d^3 x \, \sqrt{g} \, R \, \right \} \, \Psi = 0 \; ,
\eeq
with the super-Laplacian on metrics defined as
\beq
{\bf \Delta } (g) \;  \equiv  \; G_{ij,kl}  ( {\bf x} )  \;
{ \delta^2 \over \delta g_{ij} ( {\bf x} ) \, 
\delta g_{kl} ( {\bf x}  ) } \; .
\label{eq:lapl}
\eeq
We have seen before that in the discrete case one has one local
Wheeler-DeWitt equation for {\it each} tetrahedron
[see Eqs.~(\ref{eq:wd_latt}) and (\ref{eq:wd_latt1})],
which can be written as
\beq
\left \{ \, - \, ( 16 \pi \, G)^2 \; {\bf \Delta }( l^2 ) 
- \, \kappa \, \sum_{ h \subset \sigma } \delta_h \, l_h 
+ 2 \, \lambda \, V_{\sigma}  \,
\right \} \, \Psi = 0 \; ,
\label{eq:wd_3da}
\eeq
where now ${\bf \Delta }( l^2 )$ is the lattice version of the
super-Laplacian, and we have set for convenience 
$ \kappa = 2 \, n_{\sigma \, h} $.
As we shall see below, for a regular lattice of fixed coordination number,
$\kappa$ is a constant and does not depend on the location
on the lattice.
In the above expression ${\bf \Delta} ( l^2 )$ is a discretized form 
of the covariant super-Laplacian, 
acting locally on the space of $s=l^2$ variables,
\beq
{ \bf \Delta }( l^2 ) \; \equiv \; G_{ij} \; 
{\partial^2 \over \partial s_i \partial s_j}  \; ,
\label{eq:latt_lap}
\eeq
with the matrix $G^{ij}$ given explicitly in Eq.~(\ref{eq:Gij}).
Note that the curvature term involves six deficit angles
$\delta_h$, associated with the six edges of a tetrahedron.

Now, the local lattice Wheeler-DeWitt equation of Eq.~(\ref{eq:wd_3d}) applies
to a single given tetrahedron
(labeled here by $\sigma$), with one
equation to be satisfied at each tetrahedron on the lattice. 
At this point some simple additional checks can be performed.
For example, one can also construct the total Hamiltonian by
simply summing over all tetrahedra, which leads to
\beq
\left \{ - \, ( 16 \pi \, G)^2 \, \sum_{\sigma} \, {\bf \Delta } (l^2 ) 
\, + \, 2 \, \lambda \sum_{\sigma} V_{\sigma}
\, - \, \kappa \, \sum_{\sigma} \sum_{ h \subset \sigma } \, l_h \, \delta_h 
\right \} \, \Psi = 0 \; .
\label{eq:ham_tot}
\eeq
% NEW text added
The above expression represents therefore an integral over Hamiltonian 
constraints with unit density weights.
Note that indeed the second term involves the total lattice volume
(the lattice analog of $\int d^3 x \, \sqrt{g}$), and the
third one contains, as expected, the total lattice curvature
(the lattice analog of $\int d^3 x \, \sqrt{g} \, R $)\cite{reg61}.

Summing over all tetrahedra $(\sigma)$ is different from summing over
all hinges $(h)$, and the above equation is equivalent to 
\beq
\left \{ - \, ( 16 \pi \, G)^2 \,  \sum_{\sigma} {\bf \Delta } (l^2) \, + \,
2 \, \lambda \sum_{\sigma} V_\sigma \, - \,
\kappa \, q \, \sum_h \, l_h \, \delta_h \right \} \, \Psi = 0 \; ,
\eeq
where $q$ here is the lattice coordination number.
The latter is determined by how the lattice is put together (which 
vertices are neighbors to
each other, or, equivalently, by the so-called incidence matrix).
Here $q$ is therefore the number of neighboring simplexes that share a given
hinge (edge).
For a flat triangular lattice in 2d $q=6$, whereas
for the regular triangulations of $S^3$ we will be considering below
one has $q=3,\, 4,\, 5$.
For more general, irregular triangulations 
$q$ might change locally throughout the lattice.
In this case it is more meaningful to talk about an average
lattice coordination number $< \! q \! >$ \cite{itz83}.
For proper normalization in Eq.~(\ref{eq:ham_tot}) one requires 
the three-dimensional version of
Eqs.~(\ref{eq:reg_v}) and (\ref{eq:reg_r}),
which fixes the overall normalization of the curvature term
\beq
\kappa \, \equiv \, 2 \, n_{\sigma \, h} \, = \, { 2 \over q } \; ,
\label{eq:kappa}
\eeq
thus determining the relative weight of the local 
volume and curvature terms.
% NEW text added :
\footnote{
For more general, irregular triangulations $q$ might change 
locally throughout the lattice. 
Then it will be more meaningful to talk about an average 
lattice coordination number $<q>$ \cite{itz83}.}
At this point it seems worth emphasizing that from now on we 
will focus exclusively on the set of coupled {\it local} 
lattice Wheeler-DeWitt equations,  given explicitly in 
Eqs.~(\ref{eq:wd_3d}) or (\ref{eq:wd_3da}), with one 
equation for {\it each} lattice tetrahedron.

\newpage

% \vskip 40pt

\section{Choice of Coupling Constants}

\label{sec:units}

\vskip 20pt

We will find it convenient, in analogy to what is commonly done in the 
Euclidean lattice theory of gravity, 
to factor out an overall irrelevant length
scale from the problem, and set the (unscaled) cosmological constant
equal to one \cite{hw84}.
Indeed, recall that the Euclidean path integral statistical
weight always contains a factor 
$ P(V) \propto \exp (- \lambda_0 V) $, where 
$V = \int \sqrt{g} $ is the total volume on the lattice, and $ \lambda_0 $ 
is the unscaled cosmological constant.
A simple global rescaling of the metric (or edge lengths) then allows one
to entirely reabsorb this $\lambda_0$ into the local volume term.
The choice $\lambda_0=1$ then trivially fixes this overall scale once and for all.
Since $\lambda_0 = 2 \lambda / 16 \pi G $, one then has 
$\lambda = 8 \pi G  $ in this system of units.
In the following we will also find it convenient to introduce a scaled coupling 
$\tilde \lambda$ defined as
\beq
\tilde \lambda \; \equiv \; { \lambda \over 2 } 
\left ( { 1 \over 16 \pi G } \right )^2 \; .
\label{eq:tilde}
\eeq
Then for $\lambda_0=1 $ (in units of the $UV$ cutoff or, equivalently, 
in units of the fundamental lattice spacing) one has 
$\tilde \lambda = 1/ 64 \pi G $.

Two further notational simplifications will be useful in the following.
The first one is introduced in order to avoid lots of factors of 
$16 \pi$ in many of the formulas.
So from now on we shall write $G$ as a shorthand for $16 \pi\, G$,
\beq
16 \pi \, G \longrightarrow G \; .
\label{eq:shortG}
\eeq
In this new notation one has $\lambda = G / 2 $ and 
$\tilde \lambda = 1/ 4 G $.
The above notational choices then lead to a more streamlined representation of
the Wheeler-DeWitt equation, namely
\beq
\left \{  -  \, {\bf \Delta } \, + \, 
{ 1 \over  G } \, \sqrt{g} \, - \, 
{ 1 \over G^2 }  \, \sqrt{g} \; 
{}^3 \! R \, \right \} \, \Psi = 0 \;.
\label{eq:wd_3db}
\eeq
Note that we have arranged things so that now the kinetic term (the term
involving the Laplacian) has a unit coefficient.
Then in the extreme strong coupling limit ($G \rightarrow \infty$)
the kinetic term is the dominant one, followed by the volume 
(cosmological constant) term (using the facts about $\tilde \lambda$ given
above) and, finally, by the curvature term.
Consequently, at least in a first approximation, the curvature $R$ term can be
neglected compared to the other two terms, in this limit.

A second notational choice will later be dictated by the structure
of the wave function solutions, which often involve numerous
factors of $\sqrt{G}$.
It will therefore be useful to define a new coupling $g$ as
\beq
g \; \equiv \; \sqrt{G} \; ,
\label{eq:gdef}
\eeq
so that $\tilde \lambda = 4 / g^2 $ 
(the latter $g$ should not be confused with the square root of the
determinant of the metric).

\vskip 40pt

\section{Outline of the General Method of Solution}

\label{sec:method}

\vskip 20pt

The previous discussion shows that
in the strong coupling limit (large $G$) one can, at least in a
first approximation, neglect the curvature term, which will then 
be included at a later stage.
This simplifies the problem considerably, as it is the curvature 
term that introduces complicated interactions between neighboring
simplices.

Here the general procedure for finding a solution will be 
rather similar to what was
done in $2+1$ dimensions, as the formal issues in obtaining a solution
are not dramatically different.
First an exact solution is found for {\it{equilateral}} edge lengths $s$.
Later this solution is extended to
determine whether it is consistent to higher order in the weak field
expansion, where one writes for the squared edge lengths
the expansion
\beq
l_{ij}^2 \; = \; s \left( 1 \, + \, \epsilon \, h_{ij} \right) \; ,
\label{eq:wfe}
\eeq
with $\epsilon$ a small expansion parameter.
The resulting solution for the wave function can then be obtained
as a suitable power series in the $h$ variables,
combined with the standard Frobenius method, appropriate for the study of 
quantum mechanical wave equations for suitably well-behaved potentials.
In this method one first determines the correct asymptotic behavior
of the solution for small and large arguments, and later constructs
a full solution by writing the remainder as a power series or
polynomial in the relevant variable.
While this last method is rather time consuming,
we have found nevertheless that in some cases (such as the single
triangle
in $2+1$ dimensions and the single tetrahedron in $3+1$ dimensions, 
described in \cite{hw11} and also below),
one is lucky enough to find immediately an exact solution,
without having to rely in any way on the weak field expansion.

More importantly, in \cite{htw12} it was found that already in $2+1$ dimensions 
this rather laborious weak field expansion 
of the solution is not really necessary, for the following reason.
Diffeomorphism invariance (on the lattice and in the continuum) of
the theory severely restricts the form of the Wheeler-DeWitt wave function to
a function of invariants only, such as 
total three-volumes and curvatures, or powers thereof.
In other words, the wave function is found to be a function 
of invariants such as $ \int d^d x \sqrt{g} $ or
$\int d^d x \sqrt{g} \, R^n $ etc.
(these will be listed in more detail below for the specific case
of $3+1$ dimensions, where one has $d=3$ in the above expressions).

For concreteness and computational expedience,
in the following we will look at a variety of three-dimensional
simplicial lattices, including regular triangulations of the three-sphere
$S^3$ constructed as convex 4-polytopes, the latter
describing closed and connected figures
composed of lower dimensional simplices.
Here these will include the 5-cell 4-simplex or hypertetrahedron (Schl\"afli symbol $\{3,3,3\}$)
with 5 vertices, 10 edges and 5 tetrahedra;
the 16-cell hyperoctahedron (Schl\"afli symbol $\{3,3,4\}$)
with 8 vertices, 24 edges and 16 tetrahedra;
and the 600-cell hypericosahedron (Schl\"afli symbol $\{3,3,5\}$)
with 120 vertices, 720 edges and 600 tetrahedra \cite{cox48,cox74}.
Note that the Euler characteristic for all 4-polytopes that
are topological 3-spheres is zero, $\chi =N_0-N_1+N_2-N_3 =0$,
where $N_d $ is the number of simplices of dimension $d$.
We also note here that there are no known regular equilateral triangulations of the 
flat 3-torus in three dimensions, although very useful slightly irregular
(but periodic) triangulations are easily
constructed by subdividing cubes on a square lattice into 
tetrahedra \cite{hw93}.

In the following we will also recognize that there are natural sets of 
variables for displaying the results.
One of them is the scaled total volume $x$, defined as
\beq
x \; \equiv \; { 4 \, \sqrt{ 2 \lambda }  \over q \, G } 
\; \sum_{\sigma} \, V_{\sigma} \; = \;
{ 4 \, \sqrt{ 2 \lambda }  \over q \, G } \; V_{tot}  \; .
\label{eq:xdef}
\eeq
Later on we will be interested in making contact with
continuum manifolds, by taking the infinite volume 
(or thermodynamic) limit, defined in the usual way as
\bea
N_{\sigma} & \rightarrow & \infty \; ,
\nonumber \\
V_{tot} & \rightarrow & \infty \; ,
\nonumber \\
{ V_{tot} \over N_{\sigma} } & \rightarrow & {\rm const.} \; ,
\label{eq:infvol}
\eea
with $N_\sigma \equiv N_3 $ here the total number of
tetrahedra.
It should be clear that this last ratio can be used to define a 
fundamental lattice spacing $a_0$, for example via 
$ V_{tot} / N_{\sigma}  \equiv V_{\sigma} = a_0^3 / 6 \sqrt{2} $.

The full solution of the quantum mechanical problem
will, in general, require that the wave functions be
properly normalized, as in Eq.~(\ref{eq:norm}).
This will introduce at some stage wave function normalization factors
${\mathcal{N}}$,
which will later be fixed by the standard rules of quantum mechanics.
If the wave function were to depend on the total volume $V_{tot}$ only
(which is the case in $2+1$ dimensions, but not in $3+1$), 
then the relevant requirement would simply be
\beq
\lVert \Psi \rVert^2 \; \equiv \; 
\int  d \mu [g] \cdot \lvert \, \Psi [ g_{ij} ] \, \lvert^2 
\; = \;
\int_0^\infty d V_{tot} \cdot V_{tot}^m \cdot 
\lvert \, \Psi ( V_{tot} ) \, \lvert^2 \; = \; 1  \; ,
\label{eq:norm1}
\eeq
where $ d \mu [g] $ is the appropriate functional measure over the 
three-metric $g_{ij}$, and $m$ a positive real
number representing the correct entropy weighting.
But, not unexpectedly, in $3+1$ dimensions the total curvature 
also plays a role, so the above can only be regarded as 
roughly correct in the strong coupling limit (large $G$), 
where the curvature contribution to the Wheeler-DeWitt 
equation can safely be neglected.
As in nonrelativistic quantum
mechanics, the normalization condition in 
Eqs.~(\ref{eq:norm}) and (\ref{eq:norm1})
plays a crucial role in selecting out of the two 
solutions the one that is regular, and therefore satisfies 
the required wave function normalizability
condition.

To proceed further, it will be necessary to discuss 
each lattice separately in some detail.
For each lattice geometry, we will break down 
the presentation into two separate discussions.
The first part will deal with the case of no 
explicit curvature 
term in the Wheeler-DeWitt equation.
Each regular triangulation of the three-sphere
will be first analyzed separately,
and subjected to the required regularity
conditions.
Here a solution is first obtained in the
equilateral case, and later promoted on the
basis of lattice diffeomorphism invariance to
the case of arbitrary edge lengths, as was done in \cite{htw12}. 
Later a single general solution will be written down, 
involving the parameter $q$, which covers all previous
triangulation cases,
and thereby allows a first study of the infinite volume limit.
The second part deals with the extension of the 
previous solutions to the case when the curvature
term in the Wheeler-DeWitt equation is included.
This case is more challenging to treat analytically,
and the only results we have obtained so far 
deal with the large volume limit,
for which the solution is nevertheless expected
to be exact (as was the case in $2+1$ dimensions \cite{htw12}).

\vskip 40pt

\subsection{Nature of Solutions in 3+1 Dimensions}

\label{sec:sols}

\vskip 20pt

In this work we will be concerned with the solution of
the Wheeler-DeWitt equation for discrete triangulations
of the three-sphere $S^3$.
In general, for an arbitrary triangulation of a smooth
closed manifold in three dimensions, one can write down
the Euler equation 
\beq
N_0 - N_1 + N_2 - N_3 = 0
\eeq
and the Dehn-Sommerville relation
\beq
N_2 = 2 \, N_3 \; .
\eeq
The latter follows from the fact that each triangle is
shared by two tetrahedra and each tetrahedron has
four triangles, thus $2 \, N_2 = 4 \, N_3 $. 
In addition, for the regular triangulations of the three-sphere
we will be considering here, one has the additional identity
\beq
N_1 = {6 \over q} \, N_3 \; ,
\label{eq:q_def}
\eeq
where $q$ is the local coordination number, defined 
as the number of tetrahedra meeting at an edge.
For the three regular triangulations of the three-sphere
we will look at one has $q = 3,4,5 $.
The above relations then allow us to relate the number of sites
($N_0$) to the number of tetrahedra ($N_3$),
\beq
N_0 = N_3 \, \left( {6 \over q}  - 1 \right)  \; .
\label{eq:n0-n3}
\eeq
It will also turn out to be convenient to collect here a number
of useful definitions, results and identities that
apply to the regular triangulations
of the three-sphere, valid strictly when all edge lengths
take on the same identical value $l=\sqrt{s}$.
For the total volume 
\beq
V_{tot} \equiv \; \sum_{\sigma} V_{\sigma}  
\;\;  \longleftrightarrow \;\; \int d^3 x \, \sqrt{g} 
\label{eq:vtot}
\eeq
one has
\beq
V_{tot} \, = \, N_3 \, V_\sigma =  {s^{3/2} \over 6 \sqrt{2} } \; N_3 \; ,
\label{eq:vtot_q}
\eeq
whereas the total curvature
\beq
R_{tot} \; \equiv \; 2 \; \sum_h \delta_h \, l_h  
\;\;  \longleftrightarrow \;\;  \int d^3 x \, \sqrt{g} \, R 
\label{eq:rtot}
\eeq
is given by
\beq
R_{tot} = {12 \, \sqrt{s}  \over q}  \, 
\left [ 2 \, \pi - q \, \cos^{-1} \left ( \third \right )  \right ] \, N_3
\; .
\label{eq:rtot_q}
\eeq
The latter relationship can be inverted to give the parameter 
$q$ as a function of the curvature
\beq
q \; = \; q_0 \, \left( 1 - { R_{tot} \over R_{tot} 
+ {24 \, \pi \, \sqrt{s}  \over q_0}  \;  N_3  } \right)  \; ,
\label{eq:q}
\eeq
and its inverse as
\beq
q^{-1} \; = \; q_0^{-1}  + {R_{tot} \over  24 \, \pi \, \sqrt{s} \;
  N_3}  \; ,
\label{eq:qi}
\eeq
so that this last quantity is just linear in $R_{tot}$.
A very special value for $q$ corresponds to the choice
$ q= q_0 $ for which $R_{tot} =0$.
For this case one has
\beq
q_0 \; \equiv \; 
{ 2 \, \pi \over \cos^{-1} ( \third ) } = 5.1043   \; .
\label{eq:q0}
\eeq
We emphasize here again that the relationships just given above apply
to the rather special case of an equilateral triangulation.

Then, summarizing all the previous discussions, the
discretized Wheeler-DeWitt equation one wants to solve here in the
most general case is the one given in Eqs.~(\ref{eq:wd_3d}) 
or (\ref{eq:wd_3da}),
\beq
\left \{ \, -  \,  G^2 \sum_{ i,j \subset \sigma}
\, G_{ij} \, ( \sigma ) \, 
{ \partial^2 \over \partial l^2_{i} \, \partial l^2_{j}  }
\; - \; \kappa \; \sum_{ h \subset \sigma} \, l_h \, \delta_h 
\; + \; 2 \lambda  \; V_\sigma  \,
\right \} \; 
\psi [ \, l^2 \, ] \; = \; 0 \;\; ,
\label{eq:wd_3dc}
\eeq
with parameter $\kappa$ given by
\beq
\kappa = { 2 \over q} \;\; .
\eeq
Note that Eq.~(\ref{eq:wd_3dc}) still represents one equation {\it per lattice
tetrahedron}.
Thus if the lattice is made up of $N_3$ tetrahedra, the problem will
in general still require the solution of $N_3$ coupled equations
of the type given in Eq.~(\ref{eq:wd_3dc}), involving in the most general
case $N_1$ edge lengths.
As will be discussed further below, the proposed method of solution will be
quite similar to what was used earlier in $2+1$ dimensions
\cite{htw12},
namely a combination of the weak field expansion and the Frobenius
method,
which in \cite{htw12} gave the exact solution for the wavefunction
for each lattice in the limit of large areas.
If the reader is not interested here in the details of the solution
for each individual lattice, then (s)he can skip the following
sections and move on directly to Sec.~(\ref{sec:res}).

\subsection{1-Cell Complex (Single Tetrahedron)}

\label{sec:tet}

As a first case we consider here the quantum-mechanical problem
of a single tetrahedron.
One has $N_0 = 4$, $N_1 = 6$, $N_2 = 4$, $ N_3 = 1$ and
$ q = 1 $ [note that these do not satisfy the Euler and 
Dehn-Sommerville relations; only the relation between $N_1$ , 
$N_3$, and $q$, Eq.~(\ref{eq:n0-n3}), 
is satisfied for a single tetrahedron].
The single tetrahedron problem is relevant for the strong coupling
(large $G$) limit.
In this limit one can neglect the curvature term, which couples
different tetrahedra to each other, and one is left with
the local degrees of freedom, involving a single tetrahedron.

The Wheeler-DeWitt equation for a single tetrahedron
with a constant curvature density term $R$ reads
\beq
\left \{ \, - \, (16 \pi G)^2 \, G_{ij} {\partial^2 \over
\partial s_i \partial s_j} 
\;  + \; (2 \lambda - R) \, V \right \} \Psi [ \, s \, ] \;
= \; 0 \; ,
\label{eq:wd_tet}
\eeq
where now the squared edge lengths $s_1 \dots s_6 $ are all part of the
same tetrahedron, and $G_{ij}$ is given by a rather complicated, but explicit,
$ 6 \times 6 $ matrix given earlier.

As in the $2+1$ case previously discussed in \cite{htw12}, here too it is
found that, when acting on functions of the tetrahedron volume,
the Laplacian term still returns some other function of the volume only,
which makes it possible to readily obtain a full solution for the wave function.
In terms of the volume of the tetrahedron $V_\sigma$ one has the
equivalent equation for $\Psi [s]= \Psi(V_\sigma)$
(note that we have now replaced for notational 
convenience $16 \pi G \rightarrow G$)
\beq
\psi^{\prime \, \prime} \left(V_\sigma \right) +  {7 \over  V_\sigma}
\,  
\psi^{\prime} \left(V_\sigma \right)  +  {32 \, \lambda \over G^2 } \, 
\psi\left(V_\sigma \right)= 0  \; ,
\label{eq:tet_ode}
\eeq
with primes indicating derivatives with respect to $V_\sigma$.
From now on we will set the constant curvature density $R$=0.
If one introduces the dimensionless (scaled volume) variable 
\beq
x \; \equiv \; { 4 \sqrt{2 \, \lambda} \over G } \, V_{tot}  \; ,
\eeq
where $V_{tot} \equiv V_\sigma $ is the volume
of the tetrahedron,
then the differential equation for a single tetrahedron becomes simply
\beq
\psi^{\prime \, \prime} \left(x \right) + {7 \over x}  \,
\psi^{\prime} \left(x \right) 
+  \psi \left(x \right)  \; = \; 0 \; .
\label{eq:tet_ode1}
\eeq
Solutions to Eqs.~(\ref{eq:tet_ode}) or
(\ref{eq:tet_ode1}) are Bessel functions $J_m$ or $Y_m$ with  $m=3$,
\beq
\psi_R ( V_{tot} ) \; = \; {\rm const. } \; 
J_3 \left ( { 4 \sqrt{ 2 \lambda} \over G }  V_{tot} \right )
/ V_{tot}^3  \; ,
\eeq
or
\beq
\psi_S ( V_{tot} ) \; = \; {\rm const. } \; 
Y_3 \left ( { 4 \sqrt{ 2 \lambda} \over G }  V_{tot} \right ) / V_{tot}^3 \; .
\eeq
Only $J_m (x)$ is regular as $x \rightarrow 0$, 
$ J_m (x) \sim \Gamma (m+1)^{-1} (x/2)^m $.
In terms of the variable $x$ the regular solution is therefore
\beq
\psi \left( V_{tot}\right) 
\; \propto \; { J_{3} \left( x \right) \over x^{3}} 
\; \propto \; { J_{3} \left( { 4 \sqrt{ 2 \lambda} \over  G} \, 
V_{tot} \right) \over V_{tot}^{\;\;\;{3}}} \; ,
\label{eq:tet_sol}
\eeq
and the only physically acceptable wave function is
\beq
\Psi (a,b, \dots f ) \; = \; \Psi ( V_{tot} ) \; = \; {\cal N } \; 
{ J_3 \left ( { 4 \sqrt{ 2 \lambda} \over G }  V_{tot} \right ) 
\over  V_{tot}^3 }  \; ,
% \label{eq:tet_sol}
\eeq
with normalization constant
\beq
{\cal N} \; = \; 
{ 45 \sqrt{77 \pi} \over 1024 \, 2^{3/4} } \; 
\left ( { G \over \sqrt{\lambda} } \right )^{5/2} \; .
\eeq
The latter is obtained from the wave function normalization requirement
\beq
\int_0^\infty d V_{tot} \; | \, \Psi ( V_{tot} ) \, |^2 \; = \; 1 \; .
\eeq
Note that the solution given in Eq.~(\ref{eq:tet_sol}) is exact,
and a function of the volume of the tetrahedron only;
its only dependence on the values of the edge lengths of
the tetrahedron [or, equivalently, on the metric, see 
Eq.~(\ref{eq:latmet})] is through the {\it total} volume.
% NEW text added :
It is worth stressing here that in order to find the exact solution
for the wave function it would have been enough to in fact 
just consider the equilateral case. The complete solution would then
be read off immediately from this special case, if one were to
assume (as one should) that the exact wave functional is expected
to be a function of invariants only, and therefore gauge independent.

One can compute the average volume of the single tetrahedron, which
is given by
\beq
\langle \; V_{tot} \; \rangle \; \equiv \; \int_0^\infty d V_{tot} \cdot 
V_{tot} \cdot | \, \Psi ( V_{tot} ) \, |^2
\; = \; {31185 \, \pi \, G \over 262144 \, \sqrt{2 \, \lambda}  }
\; = \; 0.2643 \; {G \over \sqrt{\lambda}  } \; .
\label{eq:vol_tet}
\eeq
This last result allows us to define an average lattice spacing,
by comparing it to the value for an equilateral tetrahedron
for which $V_{tot} = (1/ 6 \sqrt{2} ) \, a_0^3 $.
One obtains
\beq
a_0 \; = \; 1.3089 \; \left ( { G \over \sqrt{\lambda}  } \right )^{1/3} \; . 
\label{eq:a0_tet}
\eeq
In terms of the parameter $\tilde \lambda$ defined in Eq.~(\ref{eq:tilde})
one has $\sqrt{\lambda}/G = \sqrt{ 2 \, \tilde \lambda}$.
With the notation of Eq.~(\ref{eq:gdef}) one has as well 
$G / \sqrt{\lambda} = \sqrt{ 2 \, G} = \sqrt{2} \, g $. 
Then for a single tetrahedron one has
$ \langle \; V_{tot} \; \rangle \; 
\equiv \langle \; V_{\sigma} \; \rangle \; = \; 0.3738 \, g $.

The single tetrahedron problem is clearly quite relevant for the limit 
of strong gravitational coupling, $1/G \rightarrow 0$.
In this limit lattice quantum gravity has a finite correlation
length, comparable to one lattice spacing,
\beq
\xi \; \sim a_0 \; .
\eeq
This last result is seen here simply as a reflection of the fact that for large $G$
the edge lengths, and therefore the metric, fluctuate more or less 
independently in different spatial regions, due to the absence of the 
curvature term in the Wheeler-DeWitt equation.
This is of course true also in the Euclidean lattice theory, 
in the same limit \cite{hw84}.
It is the inclusion of the curvature term that later leads 
to a coupling between fluctuations in different spatial regions,
an essential ingredient of the full theory.

\subsection{5-cell Complex (Configuration of 5 Tetrahedra)}

\label{sec:5cell}

The first regular triangulation of $S^3$ we will consider
is the 5-cell complex, sometimes referred to as the 
hypertetrahedron.
Here one has $N_0 =5 $, $N_1 = 10 $,
$ N_2 = 10 $, $ N_3 = 5 $ and $ q = 3 $, since there are
three tetrahedra meeting on each edge.
Then for the parameter $\kappa$ appearing in 
Eq.~(\ref{eq:wd_3dc}) one has
\beq
\kappa \, = \,   {2 \over 3} \; .
\eeq
First we will consider the case of no curvature term
in the lattice Wheeler-DeWitt equation of
Eq.~(\ref{eq:wd_3dc}).
The curvature term will be re-introduced at a later stage
[see Sec.~(\ref{sec:curv})],
as its presence considerably complicates the solution of the
lattice equations.

Solving the lattice equations directly (by brute force, one might
say) in terms of the edge length variables is a rather difficult 
task, since many edge lengths are involved,
increasingly more so for finer triangulations.
Nevertheless it can be done, to some extent, 
in $2+1$ dimensions \cite{htw12}, and possibly even in 
$3+1$ dimensions, analytically for some special cases, or numerically
for more general cases.
To obtain a full solution to the lattice equations we 
rely here instead on a simpler procedure,
already employed successfully (and checked explicitly) in $2+1$
dimensions.

% NEW paragraph expanded

First, an exact wave function solution to the lattice Wheeler-DeWitt
equations 
is obtained for the equilateral case, where all edges in the
simplicial complex 
are assumed to have the 
same length. 
This is achieved (as in \cite{htw12}) by utilizing a combination of the weak field expansion of
Eq.~(\ref{eq:wfe}) and the Frobenius (or power series expansion method) in order to 
obtain a solution to Eq.~(\ref{eq:wd_3dc}).
In order to obtain such a solution one first looks at the limit of large and small 
volumes, from which the asymptotic behavior of the solution is determined.
Note that one has one Wheeler-DeWitt equation per lattice tetrahedron, which implies
that one is seeking a solution to $N_3$ coupled equations, involving a single wave function
whose arguments are the $N_1$ edge lengths.
Nevertheless, since one is dealing here with a regular triangulation of the sphere,
all equations will have exactly the same form due to the symmetry of the problem.
It will therefore be adequate, because of this symmetry, to focus on a given single tetrahedron
and on how the associated local lattice Wheeler-DeWitt operator acts on the total wave function.
As stated previously, the latter will in general involve {\it all} lattice edge lengths.
But a further simplification arises because of the {\it locality} of
the lattice Wheeler-DeWitt equation,
which restricts interactions to edge lengths that are not too far apart.
As a consequence, when determining the structure of the wave function solution it will be 
adequate to only consider terms (local volume contributions, for example) that involve edges which 
are directly affected by the derivative terms in the local Wheeler-DeWitt operator of 
Eqs.~(\ref{eq:wd_3da}) and (\ref{eq:wd_3dc}).
Nevertheless the problem is, in spite of the above-mentioned simplifications, still of
considerable algebraic complexity in view of the many edges that still are affected
by the action of the local Wheeler-DeWitt operator. 
These generally include all the edges within the given tetrahedron, as well as a rather 
considerable number of edges located in the neighboring tetrahedra.
For a given candidate solution (written in terms of invariants, such as
the total volume and the curvature) the task is then to determine if such a solution
indeed satisfies the local Wheeler-DeWitt equation, meaning that the r.h.s. of 
Eq.~(\ref{eq:wd_3dc}) can be made to vanish, for example by a suitable
choice of wave function parameters.
Again this can be a challenging task (due to the large number of variables involved), 
unless further simplifications are invoked in order to reduce the complexity of the problem.
An additional step at this stage is therefore to constraint the solution by expanding
the r.h.s. of the local lattice Wheeler-DeWitt equation [Eq.~(\ref{eq:wd_3dc})] 
according to the weak field expansion of Eq.~(\ref{eq:wfe}).

Then, in the next step, the diffeomorphism invariance of the 
simplicial lattice theory is used to promote the previously
obtained expression for the wave function to its presumably unique
general coordinate invariant form, involving various geometric
volume and curvature terms.
It is a non-trivial consequence of the invariance properties
of the theory that such an invariant expression can be
obtained, without any further ambiguity, at least in some suitable limits
to be discussed further below (essentially, the large volume and small curvature limit).
Note that as a result of this procedure the wave function is 
ultimately {\it not} necessarily assumed to
depend on a single, global mode; instead it is still regarded as a function of all lattice metric 
degrees of freedom, as will be discussed, and used, further below (see for ex. the 
expressions given later in Eqs.~( \ref{eq:psi_largev}) and (\ref{eq:wave_asy_r}).
In a number of instances such a procedure
can be checked explicitly
and systematically within the framework of the weak field 
expansion, and used to show that the form of the relevant 
wave function solution is indeed, as expected, strongly 
constrained by diffeomorphism invariance \cite{htw12}.
In this respect the procedure we will follow here is quite different from the one used
for minisuperspace models, where the infinitely many metric degrees of freedom of
the continuum are condensed, from the very beginning and therefore already in the original
Wheeler-DeWitt equation, to one or two single modes, such as the scale factor and the
vacuum expectation of a scalar field.
% NEW footnote added
\footnote{
We should recall that in $2+1$ dimensions an exact wave functional 
was obtained for the three regular 
triangulations of the sphere (the tetrahedron, octahedron and
icosahedrons), for
arbitrary edge length assignments, in addition to the other two cases of a 
single triangle and of a regularly triangulated two-torus. 
In all the above instances it was found that the exact wave function 
solution could be described by a {\it single} function of the total area, 
of the Bessel type for strong coupling and of the confluent 
hypergeometric type in the more general case \cite{htw12}. 
As is the case here in $3+1$ dimensions, the Bessel function 
index $n$ there was found to be linearly related to the total 
number of lattice triangles $N_2$.}

In the case of the 5-cell complex, and for now without an 
explicit curvature term in the Wheeler-DeWitt equation,
one obtains the following differential equation 
\beq
\psi^{\prime \, \prime} \left(V_{tot}\right) 
+  {95 \over 9 \, V_{tot} } \,  
\psi^{\prime} \left(V_{tot} \right) 
+  {32 \, \lambda \over 9 \, G^2 } \, 
\psi\left(V_{tot} \right) \, = \, 0
\label{eq:5cell_ode}
\eeq
for a wave function that, for now, depends only on the total
volume, $ \psi = \psi \left( V_{tot} \right) $.
To obtain this result, it is assumed at first that the
simplicial complex is built out of equilateral tetrahedra;
in accordance with the previous discussion, this constraint
will be removed below.
In terms of the dimensionless variable $x$ defined as
\beq
x \, \equiv \, {4 \, \sqrt{ 2 \lambda} \over 3 \, G} \, V_{tot}
\eeq
one has the equivalent form for Eq.~(\ref{eq:5cell_ode})
\beq
\psi^{\prime \, \prime} \left(x \right)+ {95 \over 9 \, x} \,  
\psi^{\prime} \left(x \right)  + \psi \left(x \right) \,  = \, 0 \; .
\label{eq:5cell_ode1}
\eeq
This last equation can then be solved immediately, 
and the solution is
\beq
\psi \left(V_{tot}\right) 
\; \propto \;  { J_{43 \over 9} \left( x \right) \over x^{43\over 9}} 
\; \propto \;  { J_{43 \over 9} \left( {4 \, \sqrt{2 \, \lambda} \over 3 \, G} 
\, V_{tot} \right) \over V_{tot}^{\;\;\;{43\over 9}}}  \; ,
\label{eq:5cell_sol}
\eeq
up to an overall wave function normalization constant.
As in the previously discussed tetrahedron case, and also
as in $ 2+1$ dimensions, one discards the Bessel function 
of the second kind ($Y$) solution, since it is singular at the origin.

\subsection{16-cell Complex (Configuration of 16 Tetrahedra)}

\label{sec:16cell}

The next regular triangulation of $S^3$ we will consider 
is the 16-cell complex, sometimes referred to as the hyperoctahedron.
One has in this case $ N_0 =  8 $, $ N_1 = 24 $, $ N_2  = 32 $,
$ N_3  = 16 $ and $ q =  4 $, since there are four 
tetrahedra meeting on each edge.
For the parameter $\kappa$ in Eq.~(\ref{eq:wd_3dc}) one has
\beq
\kappa  \, = \, {2 \over 4} \; .
\eeq
In the case of the 16-cell complex (again for now without an
explicit curvature term in the Wheeler-DeWitt equation)
one obtains the following differential equation
\beq
\psi^{\prime \, \prime} \left(V_{tot}\right) +  
{47 \over 2 \, V_{tot}} \,  \psi^{\prime} \left(V_{tot} \right)  
+  {2 \, \lambda \over G^2 } \, \psi\left(V_{tot} \right)= 0
\label{eq:16cell_ode}
\eeq
for a wave function that depends only on the total
volume, $ \psi = \psi \left( V_{tot} \right) $.
In terms of the variable 
\beq
x \, \equiv \, { \sqrt{ 2 \, \lambda} \over G } \, V_{tot}
\eeq
one has an equivalent form for Eq.~(\ref{eq:16cell_ode})
\beq
\psi^{\prime \, \prime} \left(x \right)  
+ {47 \over 2 \, x} \,  \psi^{\prime} \left(x \right)  +  \psi \left(x
\right) \,  = \, 0 \; .
\label{eq:16cell_ode1}
\eeq
The correct wave function solution is now
\beq
\psi \left(V_{tot}\right) 
\; \propto \; { J_{45 \over 4} \left( x \right) 
\over x^{45\over 4}} 
\; \propto \; { J_{45 \over 4} 
\left( {\sqrt{ 2 \, \lambda} \over G} \, V_{tot} \right) 
\over V_{tot}^{\;\;\;{45\over 4}}} \; ,
\label{eq:16cell_sol}
\eeq
up to an overall wave function normalization constant.
Again, we discarded the Bessel 
function of the second kind ($Y$) solution, 
since it is singular at the origin.

\subsection{600-cell Complex (Configuration of 600 Tetrahedra)}

\label{sec:600cell}

The last, and densest, regular triangulation of $S^3$ we will consider here
is the 600-cell complex, often called the hypericosahedron.
For this lattice one has $ N_0  = 120 $, $ N_1 = 720 $,
$ N_2  = 1200 $, $ N_3  =  600 $ and $ q =  5$,
since there are now five tetrahedra meeting at each edge.
For the parameter $\kappa$ in Eq.~(\ref{eq:wd_3dc}) one has
\beq
\kappa  \, = \,  {2 \over 5} \; .
\eeq
For this 600-cell complex (again for now without an
explicit curvature term in the Wheeler-DeWitt equation)
one obtains the following differential equation
\beq
\psi^{\prime \, \prime} \left(V_{tot}\right) 
+  {672 \over V_{tot}} \,  \psi^{\prime} \left(V_{tot} \right)
+  {32 \, \lambda \over 25 \, G^2 } \, \psi\left(V_{tot} \right) 
\, = \, 0
\label{eq:600cell_ode}
\eeq
for a wave function that depends only on the total
volume, $ \psi = \psi \left( V_{tot} \right) $.
In terms of the variable
\beq
x \, \equiv \, {4 \, \sqrt{2 \, \lambda} \over 5 \, G} \, V_{tot}
\eeq
one has an equivalent form for Eq.~(\ref{eq:600cell_ode})
\beq
\psi^{\prime \, \prime} \left(x \right) 
+ {672  \over x} \, \psi^{\prime} \left(x \right)  
+ \psi \left(x \right)  \, = \, 0 \; .
\label{eq:600cell_ode1}
\eeq
Then the solution of the Wheeler DeWitt equation without 
a curvature term is
\beq
\psi \left(V_{tot}\right) 
\; \propto \; { J_{671 \over 2} \left( x \right) \over x^{671 \over 2}} 
\; \propto \; { J_{671 \over 2} 
\left( { 4 \, \sqrt{ 2 \, \lambda} \over  5 \, G} 
\, V_{tot} \right) \over V_{tot}^{\;\;\;{671\over 2}}}  \; ,
\label{eq:600cell_sol}
\eeq
again up to an overall wave function normalization constant.
As in previous cases, we discard the Bessel function of
the second kind ($Y$) solution, since it is singular at the origin.

\subsection{Summary and General Case for Zero Curvature}

\label{sec:res}

In this section we summarize and extend the previous results
for the wave functions, obtained so far for
the three separate cases of the 5-cell, 16-cell
and 600-cell triangulation of the three-sphere $S^3$.
The single tetrahedron case is somewhat special (it cannot
contain a curvature term), and will be left aside for
the moment.
Also, all the previous results so far apply to the
case of no explicit curvature term in the
Wheeler-DeWitt equation of Eq.~(\ref{eq:wd_3dc});
the inclusion of the curvature term will be discussed later. 
Consequently the following discussion still focuses
on the strong coupling limit, $G \rightarrow \infty $.

For the following discussion the relevant Wheeler-DeWitt 
equation is the one in Eq.~(\ref{eq:wd_3dc}),
\beq
\left \{ \, -  \,  G^2 \sum_{ i,j \subset \sigma}
\, G_{ij} \, ( \sigma ) \, 
{ \partial^2 \over \partial l^2_{i} \, \partial l^2_{j}  }
\; - \; \kappa \; \sum_{ h \subset \sigma} \, l_h \, \delta_h 
\; + \; 2 \lambda  \; V_\sigma  \,
\right \} \; 
\psi [ \, l^2 \, ] \; = \; 0 \;\; ,
% \label{eq:wdw_3c}
\eeq
which depends on the parameter
\beq
\kappa = { 2 \over q} \; ,
\eeq
where $q$ represents the number of tetrahedra meeting at an edge.
The above equation is quite general and not approximate in any way.
Nevertheless it depends on the local lattice coordination number $q$
(how the edges are connected to each other,
or, in other words, on the incidence matrix).

Now, all previous differential equations for the wave
function as a function of the total volume $V_{tot}$ 
[Eqs.~(\ref{eq:5cell_ode}), (\ref{eq:16cell_ode}) 
and (\ref{eq:600cell_ode})] can be summarized as a single
equation
\beq
\psi^{\prime \, \prime} \left( V_{tot} \right) 
+ { ( 11 + 9 \, q )  \over 2 \, q^2 } 
\, { N_3 \over V_{tot} } \;  
\psi^{\prime} \left( V_{tot} \right)  
+ {32 \over q^2} { \lambda \over  G^2} 
\; \psi \left( V_{tot} \right)  = 0 \;\; .
\label{eq:n_ode}
\eeq
Equivalently, in terms of the scaled volume variable defined as 
\beq
x \, \equiv \, 
{4 \, \sqrt{2 \, \lambda} \over q \, G} \, V_{tot} \;\; ,
\eeq
one can summarize the results of
Eqs.~(\ref{eq:5cell_ode1}), (\ref{eq:16cell_ode1}) and
(\ref{eq:600cell_ode1}) through the single equation
\beq
\psi^{\prime \, \prime} \left(x \right) 
+ { ( 11 + 9 \, q )  \over 2 \, q^2 } \, 
{N_3 \over x} \, \psi^{\prime} \left(x \right) 
+ \psi \left(x \right) \, = \, 0 \; .
\label{eq:n_ode1}
\eeq
It will be convenient here to define the (Bessel function)
index $n$ as
\beq
n  \, \equiv  \, 
{ 11 + 9 \, q \over 4 \, q^2 } \, N_3 \, - \, { 1 \over 2}  \; ,
\label{eq:n_def}
\eeq
so that for the 5-cell, 16-cell and 600-cell one has
\bea
2 \, n + 1  \, & = & { 95 \over 9 } \; \left ( q = 3, \; N_3 = 5 \right), 
\nonumber \\
& = & { 47 \over 2 } \; \left ( q = 4, \; N_3 = 16 \right), 
\nonumber \\
& = &   672 \left ( q = 5, \; N_3 = 600 \right) \; ,
\eea
respectively, and in the general case
\beq
2 \, n + 1  =   { ( 11 + 9 \, q ) \over 2 \, q^2 }  \, N_3 \; ,
\eeq
thus reproducing $n=43/9$, $45/4$ and $671/2$, respectively, 
in the three cases.
Then Eq.~(\ref{eq:n_ode1}) is just
\beq
\psi^{\prime \, \prime} \left(x \right) 
+ { 2 \, n + 1  \over x}  \;  \psi^{\prime} \left(x \right)  
+ \psi \left(x \right) \,  = \, 0 \; .
\label{eq:n_ode2}
\eeq
Consequently the wave function solutions are 
\beq
\psi \; \propto \; { J_n \left( x \right) \over x^n } \;  
\propto { J_n \left( 
{4 \, \sqrt{2 \, \lambda} \over q \, G} \, V_{tot} 
\right) 
\over 
\left ( {4 \, \sqrt{2 \, \lambda} \over q \, G} \, V_{tot} \right )^n } \; ,
\label{eq:n_sol}
\eeq
up to an overall wave function normalization constant,
thus summarizing all the results so far for the individual regular
triangulations [Eqs.~(\ref{eq:5cell_sol}), (\ref{eq:16cell_sol})
and (\ref{eq:600cell_sol})].
A more explicit, but less transparent, form for the wave
function solution is
\beq
\psi \left( V_{tot}\right) \, = \, {\cal N} \cdot
V_{tot}^ { {1\over 2} - {N_3 \left( 11 + 9 q \right) \over 4 \; q^2 }} 
\cdot
J_ {- \; {1 \over 2} + {N_3 \left(11 + 9 q \right) \over 4 q \;  q^2 }} 
\left( { 4 \, \sqrt{2 \, \lambda} \over q \; G} \; V_{tot}\right) \; ,
\label{eq:n_sol1}
\eeq
with ${\cal N}$ an overall wave function normalization constant.
Its large volume behavior is
completely determined by the asymptotic expansion of the 
Bessel $J$ function,
\beq
\psi (x) \, \simeq \, { J_n (x) \over x^n }
\; \mathrel{\mathop\sim_{ x \; \rightarrow \; \infty } } \; 
x^{-n} \, \sqrt{ 2 \over \pi x } \;
\sin \left ( x + { \pi \over 4 } - { n \, \pi \over 2 } \right )
\, + \, {\cal O} \left ( { 1 \over x^{n + {3 \over 2 }} } \right ) 
\; .
\eeq
It is also easy to see that the argument of the Bessel function 
solution $J$ in Eqs.~(\ref{eq:n_sol}) and (\ref{eq:n_sol1}) 
has the following expansion for large volumes
\beq
x \, = \, {4 \, \sqrt{2 \, \lambda} \over q_0 \, G} \, V_{tot} 
\, + \, 
{a_0^2 \over 36 \, \sqrt{2} \, \pi } \, { \sqrt{ 2 \lambda } \over G }
\; R_{tot} \; ,
\label{eq:x_strong}
\eeq
with $a_0$ ($a_0^3 \equiv 6 \sqrt{2} \, V / N_3 $) 
representing here the average lattice spacing.
Thus the second correction is of order $ (V / N_3)^{2/3} \, R_{tot} $.
Note that nothing particularly interesting is happening
in the structure of the wave function so far.
Similarly, the index $n$ of the Bessel function solution
in Eqs.~(\ref{eq:n_sol}) and (\ref{eq:n_sol1})
has the following expansion for large volumes and small curvatures,
\beq
n \, = \,
{ ( 11 + 9 \, q_0 ) \over 4 \, q_0^2 } \, N_3 - { 1 \over 2}
\, + \, 
{ ( 22 + 9 \, q_0 ) \over 96 \, \pi \, q_0 \, a_0 } \;
R_{tot}
\, + \, {\cal O} \left ( R^2 \right ) \; ,
\label{eq:n_strong}
\eeq
with $a_0$ again defined as above.
Note here that the second correction is of order $ (N_3 / V)^{1/3} \, R_{tot} $. 
It follows that the asymptotic behavior for the exponent
of the fundamental wave function solutions for large volume and 
small curvature is given by
\bea
& \pm & i \left [ \;  
{ 4 \, \sqrt{2 \, \lambda}  \over q_0 \, G} \, V_{tot} 
\; + \;  { a_0^2 \over 36 \, \sqrt{2} \, \pi } \, 
{ \sqrt{2 \, \lambda} \over G } \, R_{tot} 
\, + \, {\cal O} \left ( R^2 \right )
\; \right ]
\nonumber \\
&&
\;\;\;\;\;\;\;\;\;\;
\;\;\;\;\;\;\;\;\;\;
\;\;\;\;\;\;\;\;\;\;
\; - \; \left [ \, { 11 + 9 \, q_0 \over 4 \, q_0^2 } \, N_3   
\,+ \, { 22 + 9 \, q_0 \over \, 96 \, \pi \, q_0 \, a_0 } \, R_{tot} 
\, + \, {\cal O} \left ( R^2 \right ) \, 
\right ]\, \ln V_{tot} \; .
\label{eq:wave_asy_0}
\eea
Let us make here some additional comments.
One might wonder what concrete lattices correspond to values of $n$ greater
that $671/2$, which is after all the highest value attained for a regular
triangulation of the three-sphere, namely the 600-cell
complex.
For each of the three regular triangulations of $S^3$ with $N_0$ sites 
one has for the number of edges $N_1 = { 6 \over 6 - q } N_0 $,
for the number of triangles $N_2 = {2 \, q \over 6 - q } N_0 $
and for the number of tetrahedra $N_3 = { q \over 6 - q } N_0 $, 
where $q$ is the number of tetrahedra meeting at an edge 
(the local coordination number).
In the three cases examined previously $q$ was of course
an integer between three and five;
in two dimensions it is possible to have one more integer value of 
$q$ corresponding to the regularly triangulated torus, but
this is not possible here.
In any case, one always has for a given triangulation of
the three-sphere the Euler relation $N_0 -N_1 +N_2 -N_3 = 0$.
The interpretation of other, even noninteger, values of $q$ is then clear.
Additional triangulations of the three-sphere can be constructed
by considering irregular triangulations, where the parameter 
$q$ is now seen as an {\it average} coordination number.  
Of course the simplest example is what could be described as
a semiregular lattice, with $N_a$ edges with
coordination number $q_a$ and $N_b$ edges with coordination
number $q_b$, such that $N_a+N_b = N_1 $.
Various irregular and random lattices were considered 
in detail some time ago in \cite{itz83},
and we refer the reader to this work for a clear exposition
of the properties of these kind of lattices.
In the following we will assume that such constructions 
are generally possible, so that even non-integer values 
of $q$ are meaningful and are worth considering.

\section{Average Volume and Average Lattice Spacing}

\label{sec:avevol}

At this stage it will be useful to examine the question of
what values are allowed for the average volume.
The latter will be needed later on to give meaning to the
notion of an average lattice spacing.
In general the average volume is defined as
\beq
\langle \, V_{tot} \, \rangle \; \equiv \; 
{ \langle \Psi \vert \, V_{tot} \, \vert \Psi \rangle 
\over \langle \Psi \vert \Psi \rangle }
\; = \; 
{ \int  d \mu [g] \cdot V_{tot} [ g_{ij} ] \cdot
\vert \, \Psi [ g_{ij} ] \, \vert^2 
\over 
\int  d \mu [g] \cdot \vert \, \Psi [ g_{ij} ] \, \vert^2 }  \; ,
\label{eq:ave_vol}
\eeq
where $ d \mu [g] $ is the appropriate (DeWitt) 
functional measure over three-metrics $g_{ij}$.

Now consider the wave function obtained given in Eq.~(\ref{eq:n_sol}),
with $n$ defined in Eq.~(\ref{eq:n_def}).
This wave function is relevant for the strong coupling limit,
where the explicit curvature term in the Wheeler-DeWitt equation
can be neglected.
In this limit one can then compute the average total volume
\beq
\langle \, V_{tot} \, \rangle
\;  = \; {\int_0^\infty \, d V_{tot} \cdot  
V_{tot} \cdot \vert \psi\ ( V_{tot} ) \vert^2 
\over 
\int_0^\infty \, d V_{tot} \cdot
\vert \psi ( V_{tot} )\vert^2} \; .
\eeq
One then obtains immediately for the average volume 
of a tetrahedron
\beq
\langle \, V_\sigma \, \rangle 
\; = \; { 2^{-{3\over2} - 2 \, n} \; 
\Gamma \; \left( n - \half \right) 
\; \Gamma \left(2 \, n + \half \right)  \over
\Gamma\left(n\right)^3 \, N_3 } \cdot
{ q \, G \over \sqrt{\lambda} } \; .
\eeq
If the whole lattice is just a single tetrahedron,
then one has $n=3$ and consequently
\beq
\langle \, V_\sigma \, \rangle 
\; = \; {31185 \, \pi \, G \over 262144 \, \sqrt{2} \, \sqrt{\lambda}} 
\; = \;  0.2643 \, { G \over \sqrt{\lambda}} \; ,
\label{eq:vave_tet}
\eeq
from which one can define an average lattice spacing $a_0$
via $ \langle \, V_{\sigma} \, \rangle = a_0^3 / 6 \, \sqrt{2} $.
For large $N_3$ one has
\beq
a_0^3 \;  = \; {3 \, \sqrt{ 11 + 9 \, q }
\over 2 \, \sqrt{ 2\, \pi } \, N_3 } 
\, { G \over \sqrt{\lambda} } \; .
\eeq
But in general one cannot assume a trivial entropy
factor from the functional measure, and
one should evaluate instead
\beq
\langle \, V_{tot} \, \rangle
\;  = \; {\int_0^\infty \, d V_{tot}  
\cdot V_{tot}^m \cdot V_{tot} \cdot 
\vert \, \psi\ ( V_{tot} ) \, \vert^2 
\over 
\int_0^\infty \, d V_{tot} \cdot V_{tot}^m \cdot
\vert \, \psi ( V_{tot} )\, \vert^2} \; ,
\label{eq:entropy}
\eeq
with some power $ m = c_0 \, N_3 $ and $c_0$ a real 
positive constant.
One then obtains for the average volume of a single
tetrahedron
\beq
\langle \, V_\sigma \, \rangle 
\, = \, { 1 \over N_3 } \; \langle \, V_{tot} \, \rangle 
\; = \; 
\sqrt{ c_0 \left [ 11 + q_0 (9 - c_0 \, q_0 ) \right ] } \;
{ G \over 8 \, \sqrt{2 \, \lambda} } \; ,
\label{eq:vave_c0}
\eeq
which is finite as $N_3 \rightarrow \infty $.
Note that in order for the above expression to make
sense one requires $ c_0 < (11+9 q_0)/q_0^2 \, \simeq \, 2.185 $.
If the exponent in the entropy factor is too large,
the integrals diverge.
One then finds that the corresponding lattice spacing is given by
\beq
a_0^3 \; = \; 
\sqrt{ c_0 \left [ 11 + q_0 (9 - c_0 \, q_0 ) \right ] }
\; { 3 \, G \over 4 \, \sqrt{\lambda} } \; .
\label{eq:a0_ent}
\eeq
The lesson learned from this exercise is that in gravity
the lattice spacing $a_0$ (the fundamental length scale, or
the ultraviolet cutoff if one wishes) is itself
dynamical, and thus set by the bare values of
$G$ and $\lambda$.
In a system of units for which $\lambda_0=1$ one
then has $ a_0 \sim g^{1/3}$.
Either way, the choice for $a_0$ has no immediate
direct physical meaning,
and has to be viewed instead in the context
of a subsequent consistent renormalization procedure.
In the following it will be safe to assume,
based on the results of 
Eqs.~(\ref{eq:a0_tet}) and (\ref{eq:a0_ent}) that
\beq
a_0^3 \; = \; f^3 \, { G \over \sqrt{\lambda} } \; ,
\label{eq:a0_gen}
\eeq
in units of the UV cutoff, 
where $f$ is a numerical constant of order one
(for concreteness,
in the single tetrahedron case one has $f \approx 1.3089 $).

\section{Large Volume Solution for Nonzero Curvature}

\label{sec:curv}

The next task in line is to determine the form of the wave function
when the curvature term in the 
Wheeler-DeWitt equation of Eq.~(\ref{eq:wd_3dc}) is not zero.
In particular we will be interested in the changes
to the wave function given in  
in Eqs.~(\ref{eq:n_sol}) and (\ref{eq:n_sol1}), with
argument $x$ in Eq.~(\ref{eq:x_strong}) and 
parameter $n$ in Eq.~(\ref{eq:n_strong}).
We define here the total integrated curvature $ R_{tot} $ as
in Eq.~(\ref{eq:rtot}), which is of course different from the 
local curvature appearing in the lattice
Wheeler DeWitt equation of Eq.~(\ref{eq:wd_3dc}),
\beq
R_{\sigma} \equiv \sum_{ h \subset \sigma} \delta_h \, l_h \; .
\eeq
In order to establish the structure of the solutions for
large volumes $V_{tot}$ we will assume, based in part
on the results of the previous sections, and on the 
analogous calculation in $2+1$ dimensions \cite{htw12},
that the fundamental wave function solutions for large 
volumes have the form
\beq
\exp \left \{ \pm \, i \left( \alpha \int d^3 x \sqrt{g}\, 
+ \beta \int d^3 x \sqrt{g} \, R 
+ \gamma \int d^3 x \sqrt{g} \, R^2 
+ \delta \int d^3 x \sqrt{g} \, R_{\mu \nu} R^{\mu \nu} + \cdots \right)
\right \} \; .
\label{eq:psi_asym}
\eeq
Note here that the structure of the above expression, and the
nature of the terms that enter into it, are
basically dictated by the requirement 
of diffeomorphism invariance as it applies to the argument of
the wave functional.
Apart from the cosmological term, allowed terms are all the 
ones that can be constructed from the
Riemann tensor and its covariant derivatives, for a 
a fixed topology of 3-space.
Clearly, at large distances (infrared limit) the most important terms
will be the Einstein and cosmological terms, with
coefficients $\beta$ and $\alpha$, respectively.
In three dimensions the Riemann and Ricci
tensor have the same number of algebraically
independent components (6), and are related to
each other by
\beq
R^{\mu\nu}_{\;\;\;\; \lambda \sigma} \; = \; 
\epsilon^{\mu\nu\kappa} \; \epsilon_{\lambda\sigma\rho} \; 
\left ( R^\rho_{\;\;\kappa} \, - \, \half \, \delta^\rho_{\;\;\kappa}
\right ) \; .
\label{eq:riem-3d}
\eeq
The Weyl tensor vanishes identically, and one has
\beq
R _ { \mu \nu \lambda \sigma }  R ^ { \mu \nu \lambda \sigma } -
4  R _ { \mu \nu }  R ^ { \mu \nu } - 3  R ^ 2  \; = \; 0
\;\;\;\;\;\;
C _ { \mu \nu \lambda \sigma }  C ^ { \mu \nu \lambda \sigma } = 0 \; .
\eeq
As a consequence, there is in fact only {\it one} local curvature 
squared term one can write down in three 
spatial dimensions.
Nevertheless, higher derivative terms will only 
become relevant at very short
distances, comparable or smaller than the Planck
length $\sqrt{G}$; in the scaling limit it
is expectd that these can be safely neglected.

When expressed in lattice language, the above form 
translates to an ansatz of the form
\beq
\exp \left \{ \, \pm \, i \left( c_0 \, V_{tot} 
\, + \, c_1 \, R_{tot}^{\; m} \, \right) \right \}  \; ,
\eeq
with $m$ assumed to be an integer.
In addition, from the studies of lattice gravity $2 + 1$ dimensions
one expects a $ \ln V_{tot} $ term as well in the argument of the
exponential \cite{htw12}.
This suggests a slightly more general ansatz ,
\beq
\exp \left \{ \, \pm \, i \left( \; c_0 \, V_{tot} 
+ c_1 \, R_{tot}^{\; m}  \; \right) 
+ c_2 \, \ln V_{tot} + c_3 \, \ln R_{tot} \, \right \} \;\; .
\label{eq:ansatz}
\eeq
The next step is to insert the above expression 
into the lattice Wheeler-DeWitt equation 
Eq.~(\ref{eq:wd_3dc}) and determine the values 
of the five constants $c_0 \dots c_3 , \, m$.
This can be done consistently just to leading order in 
the weak field expansion of Eq.~(\ref{eq:wfe}),
which is entirely adequate here, as it will provide
enough information to uniquely determine the coefficients.
Here we will just give the result of this exercise.
For the 5-cell complex ($q=3$) one obtains
\beq
\psi 
\sim \exp \left\{ \pm \, i \left(\;  { 4 \sqrt{2} \, 
\sqrt{\lambda}  \over 3 \, G} \, V_{tot} 
\; - \;  { \sqrt{2} \over  G \, \sqrt{\lambda} } \, R_{tot} \;\right) 
- {95  \over 18} \, \ln V_{tot} \;\right\} \; ,
\label{eq:5cell_largev}
\eeq
whereas for $16$-cell complex ($q=4$) one finds
\beq
\psi 
\sim \exp \left\{ \pm \, i 
\left(\;  {  \sqrt{2} \, \sqrt{\lambda}  \over  \, G} \, V_{tot} 
\; - \;  { 3 \, \sqrt{2}  \over 4 \, G \, \sqrt{\lambda} } \, R_{tot}
\;\right)   
- {47  \over 4} \, \ln V_{tot} \;\right\} \; ,
\label{eq:16cell_largev}
\eeq
and finally for $600$-cell complex ($q=5$)
\beq
\psi 
\sim \exp \left\{ \pm \, 
i \left(\;  {  4 \, \sqrt{2} \, \sqrt{\lambda}  \over  5 \, G} \, V_{tot} 
\; - \;  { 3 \, \sqrt{2} \over 5 \, G \, \sqrt{\lambda} } \, 
R_{tot} \;\right)   - {336} \, \ln V_{tot} \;\right\} \; .
\label{eq:600cell_largev}
\eeq
These expressions allow us again to identify the answer for
general $q$ as
\beq
\psi 
\sim \exp \left \{ \pm \, 
i \left( \;  { 4 \, \sqrt{2 \, \lambda}  \over q \, G} \, V_{tot} 
\; - \;  { 3 \sqrt{2} \over q \, G \, \sqrt{\lambda} } \, R_{tot}
\right )
\, - \, { (11 + 9 \, q ) \, N_3  \over 4 \, q^2 } 
\, \ln V_{tot} \; \right \} \; .
\label{eq:psi_largev}
\eeq
Note that in deriving the above results
we considered the large volume limit $V \rightarrow \infty$,
treating the number of tetrahedra $N_3$ as a fixed parameter.
% NEW sentence added
In writing down this last result we have used the fact that such a 
$q$ dependence of the curvature 
term is expected on the basis of Eq.~(\ref{eq:kappa}) , 
and similarly for the volume term in view of 
Eq.~(\ref{eq:xdef}).  In addition, the $log$ term is expected on 
general grounds to have a coefficient proportional to the number 
of lattice tetrahedral $N_3$,  as it does (exactly) in $2+1$ 
dimensions \cite{htw12}. 
Note that later the effect of the $log$ term will be in part compensated 
by the measure (or entropy) contribution of Eq.~(\ref{eq:entropy}).
% NEW footnote added
\footnote{
A rather similar procedure was successfully used earlier in $2+1$ dimensions, where it was 
found that the three regular triangulations of the sphere, the single traingle and the
regular triangulation of the torus were all described, for large areas and to 
all orders in the weak field expansion, by a single wave functional 
involving confluent hypergeometric functions, with the total area and 
total curvature serving as arguments. 
The resulting extrapolation to the infinite volume limit yielded exact 
gravitational scaling exponents \cite{htw12} in rough agreement 
(to about 6\%) with results obtained earlier by numerical integration in 
the Euclidean lattice theory of gravity \cite{hw93}. }

Then from the previous expression we can now read off 
the values for the various coefficients, namely
\bea
c_0 & = &  {4 \, \sqrt{2 \, \lambda}  \over q \, G} 
\nonumber \\
c_1 & = &  - \;  {3 \sqrt{2} \over q \, G \, \sqrt{\lambda} } 
\nonumber \\
c_2 & = &  - \, { (11 + 9 \, q ) \, N_3 \over 4 \, q^2 }
\nonumber \\
c_3 & = & 0
\label{eq:coeffs}
\eea
with the only possible value $m = 1$.

In order to make contact with the strong coupling result
for the wave function derived in the previous sections
[Eqs.~(\ref{eq:n_sol1}), (\ref{eq:x_strong}),
(\ref{eq:n_strong}) and (\ref{eq:wave_asy_0})],
one needs to again expand the above answer for small curvatures.
One obtains for the exponent of the wave function the following
expression
\bea
\pm \, i \left \{ \;  
{ 4 \, \sqrt{2 \, \lambda}  \over q_0 \, G} \, V_{tot} 
\; + \;  \left( { a_0^2 \over 36 \, \sqrt{2} \, \pi } \, 
{ \sqrt{2 \, \lambda} \over G } 
\, - \,  {6 \over q_0 \, G \, \sqrt{2 \, \lambda} } \right )
\, R_{tot} 
\, + \, {\cal O} \left ( R^2 \right )
\; \right \}
\nonumber \\
\;\;\;\;\;\;\;\;\;\;
\;\;\;\;\;\;\;\;\;\;
\; - \; \left \{ \, { 11 + 9 \, q_0 \over 4 \, q_0^2 } \, N_3   
\,+ \, { 22 + 9 \, q_0 \over \, 96 \, \pi \, q_0 \, a_0 } \, R_{tot} 
\, + \, {\cal O} \left ( R^2 \right ) \, 
\right \} \, \ln V_{tot}  \; ,
\label{eq:wave_asy_r}
\eea
with $a_0$ again representing the average lattice spacing,
$a_0^3 \equiv 6 \sqrt{2} \, V / N_3 $.
This finally determines uniquely the coefficients 
$\alpha$ and $\beta$ appearing in Eq.~(\ref{eq:psi_asym}),
\bea
\alpha &  = & {4 \over q_0 } \cdot 
{ \sqrt{2 \, \lambda}  \over G }  
\nonumber \\
\beta & = & { a_0^2 \over 36 \, \sqrt{2} \, \pi } \cdot
{ \sqrt{2 \, \lambda}  \over G }
\, -  \;  
{ 6 \over \; q_0 } \cdot { 1 \over G \, \sqrt{2 \, \lambda} }  \; .
\label{eq:alpha}
\eea
The most important result so far is
the appearance of two contributions
of opposite sign in $\beta$,
signaling the appearance of
a critical value for $G$ where $\beta$ vanishes.

This critical point is located at 
$ \lambda_c \, = \, 108 \, \sqrt{2} \pi / q_0 \, a_0^2 $
or, in a system of units where $ \lambda = G/2 $ [see
  Sec.~(\ref{sec:units})],
% NEW footnote :
\footnote{
As in the Euclidean lattice gravity case, one does not expect the 
critical coupling $G_c$ to represent a universal quantity; 
its value will still reflect specific choices made in defining the underlying 
lattice discretization, and therefore more generally in specifying a 
suitable ultraviolet cutoff (this fact is known in field theory language 
as scheme dependence). These circumstances can be seen here already 
when looking at the simplest regular lattices enumerated previuosly in 
this work, and which are clearly not unique choices even for a fixed number of sites.
In addition, one expects a further dependence of $G_c$ on the choice 
of functional measure and therefore on the supermetric [see for ex. Eq.~(\ref{eq:vave_c0})].
In the present context this leads, for example, to a dependence of 
the results on the parameter $f$ of Eq.~(\ref{eq:a0_gen}).
Nevertheless one would expect, based largely on universality
arguments, 
that critical exponents and scaling dimensions (such as the ones 
obtained exactly in \cite{htw12}) should be universal, and therefore 
independent of the specific details of the ultraviolet cutoff, 
whose introduction nevertheless is essential at some stage in 
order to regularize the inevitable quantum infinities.}
\beq
G_c \, = \, 
{ 216 \, \sqrt{2} \pi \over q_0 } \cdot { 1 \over a_0^2 } \; .
\label{eq:G_c}
\eeq
But since the average lattice spacing $a_0$ is itself 
a function of $G$ and $\lambda$
[see Eqs.~(\ref{eq:a0_tet}), (\ref{eq:a0_ent}) and
(\ref{eq:a0_gen})]
one obtains in the same system of units
\beq
G_c \, = \, { 36 \; 2^{3/8} \, 3^{1/4} \, \pi^{3/4}
\over f^{3/2} \; q_0^{3/4} } \, \simeq \, 28.512  \; ,
\eeq
using the value of $f$ for the single tetrahedron,
or equivalently $g_c \simeq 5.340 $, a rather large value.
Nevertheless we should keep in mind that in this paper
we are also using a system of units
where we set $16 \pi G \rightarrow G $.
So, in a conventional system of units, one has the more
reasonable result $G_c \approx 0.5672 $ in units of the fundamental
UV cutoff.
\footnote{
One can compare the above value for $G_c$ obtained in the Lorentzian $3+1$ 
theory with the corresponding value in the Euclidean four-dimensional
theory.
There one finds $G_c \approx 0.6231 $ \cite{ham00}, which is within ten
percent of the above quoted value.
The two $G_c$ values are not expected to be the same in the two
formulations, due to the different nature of the UV cutoffs.
In particular, in the lattice Hamiltonian formulation the
continuum limit has already been taken in the time direction.
Nevertheless, it is encouraging that they are quite comparable in magnitude.}
Evidence for a phase transition in lattice gravity in $3+1$
dimensions was also seen earlier from an application of
the variational method, using Jastrow-Slater correlated product 
trial wavefunctions \cite{hw11}. 
Note that the results of Eqs.~(\ref{eq:wave_asy_r}) and (\ref{eq:alpha})
imply a dependence of the fundamental wave function on the curvature,
of the type
\beq
\psi ( R ) \; \sim \; e^{ \, \pm \, i \, R_{tot} / R_0 }  \; ,
\label{eq:curv_fluc}
\eeq
with $R_0$ a characteristic scale for the total, integrated curvature.
Thus $ R_0 \sim 1 / ( g - g_c ) $ with $G_c$, and therefore 
$g_c = \sqrt{G_c} $, given in Eq.~(\ref{eq:G_c}).
Therefore at the critical point fluctuations in the
curvature become unbounded, just as is the case for
the fluctuations in a scalar field when the renormalized mass
approaches zero.
\footnote{
It is tempting to try to extract a critical exponent
from the result of Eq.~(\ref{eq:curv_fluc}).
In analogy to the wave functional for a free scalar
field with mass $m$, and thus correlation length $\xi=1/m$,
one would obtain for the correlation length exponent $\nu $
(with $\nu$ defined by $\xi \sim \vert g - g_c \vert^{- \nu }$)
from the above wave function the {\it semi-classical} 
estimate $ \nu \, = \, \half $.
In the $2+\epsilon$ perturbative expansion for pure gravity one finds
in the vicinity of the UV fixed point
$\nu^{-1} = (d-2) + { 3 \over 5} (d-2)^2 + {\cal O} ( (d-2)^3 )$
\cite{wei79,eps,aid97}.
The above lowest order lattice result would then agree 
only with the leading, semi-classical term.}
% NEW text added :
Note that since at the critical point $G_c$ the curvature term
vanishes, 
further investigations there would require the retention of curvature 
squared terms, which in general are not expected to be zero. 
One would then expect, based again on invariance arguments, 
that the leading contribution there should come from the $TT$ 
mode contribution, which is indeed quadratic in the curvature. 

At this stage one can start to compare with the results
obtained previously without the explicit curvature term
in the Wheeler-DeWitt equation,
Eqs.~(\ref{eq:x_strong}) and (\ref{eq:n_strong}).
The main change is that here one would be led to identify
\beq
x \, = \, {4 \, \sqrt{2 \, \lambda} \over q_0 \, G} \, V_{tot} 
\, + \, \left (
{ a_0^2 \over 36 \, \sqrt{2} \, \pi } \cdot
{ \sqrt{2 \, \lambda}  \over G }
\, -  \;  
{ 6 \over \; q_0 } \cdot { 1 \over G \, \sqrt{2 \, \lambda} } 
\right ) \, R_{tot}  \; ,
\label{eq:x_r}
\eeq
so that the Bessel function argument $x$ 
[see Eq.~(\ref{eq:x_strong})] now contains a new contribution, 
of opposite sign, in the curvature term.
Its origin can be traced back to the new curvature 
contribution $c_1$ in Eq.~(\ref{eq:coeffs}), 
which in turn arises because of the explicit
curvature term now present in the full Wheeler-DeWitt equation.
On the other hand, as is already clear from
the result for $c_2$ in Eq.~(\ref{eq:coeffs}),
the index $n$ of the Bessel function solution
in Eqs.~(\ref{eq:n_sol}) and (\ref{eq:n_sol1})
is left unchanged,
\beq
n \, = \,
{ 11 + 9 \, q_0 \over 4 \, q_0^2 } \; N_3 - { 1 \over 2}
\, + \, 
{ 22 + 9 \, q_0 \over 96 \, \pi \, q_0 \, a_0 } \;
R_{tot}
\, + \, {\cal O} \left ( R_{tot}^2 \right )   \; ,
\label{eq:n_r}
\eeq
with again an average lattice spacing $a_0$ defined as before.

But there is a better way to derive correctly the modified
form of the wave function.
From the asymptotic solution for the wave function
of Eq.~(\ref{eq:psi_largev})
it is possible to first obtain a partial differential equation
for $\psi ( R_{tot}, V_{tot} ) $.
The equation reads (in the following we shall write
$R_{tot} $ as $R$ and $V_{tot}$ as $V$ to avoid 
unnecessary clutter)
\beq
{\partial^2 \psi \over \partial V^2}  
\, + \, c_V \, {\partial \, \psi \over \partial \, V}  
\, + \, c_R \, {\partial \, \psi \over \partial \, R}  
\, + \, c_{VR} \, {\partial^2 \, \psi \over \partial \, V
\partial \, R} 
\, + \, c_{RR} \, {\partial^2 \, \psi \over \partial \, R^2} 
\, + \, c_\lambda \, \psi 
\, + \, c_{curv} \, \psi 
\, = \, 0 \; .
\label{eq:ode_r}
\eeq
The coefficients in the above equation are given by
\bea
c_V & = & 
{11 + 9 \, q  \over 2 \; q^2} \cdot {N_3 \over V}
\, = \, 
{11 + 9 \, q_0 \over 2 \; q_0^2} \cdot {N_3 \over V}
\, + \, 
{ 22 + 9 \, q_0 \over 48 \, \sqrt{2} \, 3^{1/3} \, \pi \, q_0}
\cdot { N_3^{1/3} \, R \over V^{4/3}} \, + \, {\cal O} ( R^2 )
\nonumber \\
c_R & = & 
- { 2 \over 9 } \, { R \over V^2 }
\, + \, {11 + 9 \, q_0 \over 6 \; q_0^2} \cdot { N_3 \, R \over V^2 }
\, + \, {\cal O} ( R^2 )
\nonumber \\
c_{VR} & = & 
{ 2 \over 3 } \, { R \over V } \, + \, {\cal O} ( R^2 )
\nonumber \\
c_{RR} & = & { 2 \over 9 } \, { R^2 \over V^2 }
\nonumber \\
c_\lambda & = & 
{32 \, \lambda \over q^2 \, G^2} \, = \, 
{ 32 \over G^2 \, q_0^2 } \, + \, 
{ 4 \sqrt{2} \lambda \over 3 \, 3^{1/3} \, \pi \, q_0 \, G }
\cdot { R \over N_3^{2/3} \, V^{1/3} } \, + \, {\cal O} ( R^2 )
\nonumber \\
c_{curv} & = & 
- { 16 \over G^2 \, q^2 } \cdot { R \over V }
\, = \, - { 16 \over G^2 \, q_0^2 } \cdot { R \over V }
\, + \, {\cal O} ( R^2 ) \; .
\label{eq:ode_r_coeffs}
\eea
Note that in the small curvature, large volume limit 
[this is the limit in which,
after all, Eq.~(\ref{eq:psi_largev}) was derived] one
can safely set the coefficients $c_R$ and $c_{RR}$ to zero.
It is then easy to check that the solution
in Eq.~(\ref{eq:psi_largev}) satisfies
Eqs.~(\ref{eq:ode_r}) and (\ref{eq:ode_r_coeffs}),
up to terms of order $1/V^2$.
Also note that here, and in 
Eqs.~(\ref{eq:5cell_largev}), (\ref{eq:16cell_largev}), 
(\ref{eq:600cell_largev}) and (\ref{eq:psi_largev}),
we take the large volume limit $V \rightarrow \infty$,
treating the number of tetrahedra $N_3$ as a large, fixed parameter.
A differential equation in the variable $V$ only can be derived
as well (with coefficients that are functions of $R$),
but then one finds that the required coefficients
are not real, which makes this approach less
appealing.
% NEW footnote :
\footnote{
It would of course be of some interest to derive a result 
similar to Eq.~(\ref{eq:wave_asy_r}) using
an entirely different set of methods, such as the $WKB$
approximation. 
Such an approximation was
discussed, again on the lattice, in \cite{hw11}, but the resulting 
equations there turned out to be too complicated to solve.
In the context of a continuum $WKB$ approximation, one would 
expect the approximate results for the wave function to contain some
remnants of short distance infinities, and therefore depend, at least
in part, implicitly or explicitly, on the specifics of the ultraviolet 
regularization procedure. 
But, more generally, one would expect such a continuum 
expansion to be poorly convergent in four dimensions, in view 
of the perturbative non-renormalizability of ordinary gravity.
The lattice methods presented here are, on the other hand,
genuinely non-perturbative in nature, and therefore not immediately
affected by the escalating divergences encountered in the continuum 
treatment in four dimensions.}

In the limit $R \rightarrow 0 $
Eq.~(\ref{eq:ode_r}) reduces to
\beq
{\partial^2 \psi \over \partial V^2}  
\, + \, 
{11 + 9 \, q_0 \over 2 \; q_0^2} \cdot {N_3 \over V} 
\cdot {\partial \, \psi \over \partial \, V}  
\, + \, 
{ 32 \, \lambda \over G^2 \, q_0^2 } \, \psi 
\, = \, 0 \; ,
\label{eq:ode_r0}
\eeq
which is essentially Eq.~(\ref{eq:n_ode}) in the same 
limit, with solution given previously in Eq.~(\ref{eq:n_sol}).

\section{Nature of the Wave Function Solution ${\bf \psi}$}

\label{sec:psi}

In this section we discuss some basic physical
properties that can be extracted from the wave function 
solution $\psi (V,R)$.
So far we have not been able to find a general
solution to the fundamental Eq.~(\ref{eq:ode_r}),
but one might suspect that the solution is still close
to a Bessel or hypergeometric function, possibly with arguments
``shifted'' according to Eqs.~(\ref{eq:x_r})
and (\ref{eq:n_r}), as was the case in $2+1$ dimensions.
As a consequence, some physically motivated
approximations will be necessary in the following discussion.
Let us discuss here in detail one possible approach.
If one sets the troublesome coefficient $c_{VR} = 0$ in
Eq.~(\ref{eq:ode_r}), and keeps only the leading
term in $c_V$, then the relevant differential 
equation becomes
\beq
{\partial^2 \psi \over \partial V^2}  
\, + \, c_V \, {\partial \, \psi \over \partial \, V}  
\, + \, c_\lambda \, \psi 
\, + \, c_{curv} \, \psi 
\, = \, 0 \; ,
\label{eq:ode_rs}
\eeq
with coefficients given in Eq.~(\ref{eq:ode_r_coeffs}),
except that from now on only the leading term in $c_V$ and
$c_{\lambda}$ will be retained (otherwise it seems
again difficult to find an exact solution).
Note that the above equation still contains an excplicit
curvature term proportional to $R$, from $c_{curv}$.
Now a complete solution can be found in terms of the confluent 
hypergeometric function of the first kind, $_1 F_1 (a,b,z) $
\cite{as72,as12,nist10}.
Up to an overall wave function normalization constant, it is
\bea
\psi ( V, \, R) & \simeq &
e^{ - { 4 \, i \sqrt{2 \lambda} \, V \over q_0 \, G } }
\cdot { 
\Gamma \left ( 
{ ( 11 + 9 \, q_0 ) \, N_3 \over 4 \, q_0^2 } 
+ { i \, \sqrt{2}\, R \over q_0 \, G \, \sqrt{\lambda} }
\right )   
\over
\Gamma \left ( 
1 - { ( 11 + 9 \, q_0 ) \, N_3 \over 4 \, q_0^2 } 
+ { i \, \sqrt{2}\, R \over q_0 \, G \, \sqrt{\lambda} }
\right ) 
\, }
\nonumber \\
&&
\;\;\;\;\;\;\;\;\;\;
\times \;
_1 F_1 \left (
{ ( 11 + 9 \, q_0 ) N_3 \over 4 \, q_0^2 } 
- { i \, \sqrt{2} \, R \over q_0 \, \sqrt{\lambda} \, G }
, \;
{ ( 11 + 9 \, q_0 ) N_3 \over 2 \, q_0^2 } 
, \;
{ 8 \, i \sqrt{2 \lambda} \, V \over q_0 \, G }
\right )  \; .
\label{eq:ode_sol}
\eea
Here again $q_0$ is just a number, given previously in
Eq.~(\ref{eq:q0}),
and $N_3$ the total number of tetrahedra for a given
triangulation of the manifold.
Note that this last solution still retains three key
properies:
it is a function of geometric invariants ($V,R)$ only;
it is regular at the origin in the variable $V$ (the irregular
solution is discarded due to the normalizability constraint);
and finally it agrees, as it should, with the zero curvature solution
of Eqs.~(\ref{eq:n_sol}) and (\ref{eq:n_sol1}) in the limit $R=0$.

The above wave function exhibits some intriguing similarities with the exact
wave function solution found in $2+1$ dimensions; the
difference is that the total curvature $R$ here
plays the role of the Euler characteristic $\chi$ there.
Let us be more specific, and discuss each argument separately.
For the arguments of the confluent hypergeometric function
of the first kind, $_1 F_1 (a,b,z) $, one finds again
$ b =2 a $ for $R=0$, with both $a$ and $b$ proportional 
to the total number of lattice sites, as in $2+1$ dimensions \cite{htw12}.
Specifically, here one has
\beq
Re (a) \; = \; { 11 + 9 \, q_0 \over 4 \, q_0^2 } \, N_3 \; \approx \; 
0.5464 \, N_3 \; ,
\eeq
whereas in $2+1$ dimensions the analogous result is
\beq
Re (a) \; = \; \quarter \, N_2 \; .
\eeq
The curvature contribution in both cases
then appears as an additional contribution to the first argument 
($a$), and is purely imaginary.
Here one has
\beq
Im (a) \; = \; - \, { \sqrt{2} \over q_0 \, \sqrt{\lambda} \, G } 
\; \int d^3 x \, \sqrt{g} \, R \; ,
\eeq
whereas in $2+1$ dimensions the corresponding result is
\beq
Im (a) \; = \; - \, { 1 \over 2 \, \sqrt{2 \, \lambda} \, G } 
\; \int d^2 x \, \sqrt{g} \, R \; .
\eeq
Finally, here again the third argument $z$ is purely imaginary and
simply proportional to the total volume.
From the above solution
\beq
z \; = \; i \, { 8 \, \sqrt{2 \lambda} \over q_0 \, G } \, \int d^3 x
\, \sqrt{g} \; ,
\eeq
whereas in $2+1$ dimensions 
\beq
z \; = \; i \, { 2 \, \sqrt{2 \lambda} \over G } \, \int d^2 x \,
\sqrt{g} \; .
\eeq
Nevertheless we also find some significant differences
when compared to the exact $2+1$-dimensional result, 
most notably the various gamma-function
factors involving the curvature $R$, which are entirely absent
in the lower dimensional case, as well as the fact that 
the critical (UV fixed) point is 
located at some finite $G_c$ here [see Eq.~(\ref{eq:G_c})], 
whereas it is exactly at $G_c=0$ in $2+1$ dimensions \cite{htw12}.

Let us now continue here with a discussion of the main properties
of the wave function in Eq.~(\ref{eq:ode_sol}).
First let us introduce some additional notational simplification.
By using the coupling $g$
[see Sec.~(\ref{sec:units}) and Eq.~(\ref{eq:gdef})]
one can make the above expression for $\psi$ slightly more transparent
\bea
\psi ( V, \, R) & \simeq &
e^{ - { 4 \, i \, V \over q_0 \, g } }
\cdot { 
\Gamma \left ( 
{ ( 11 + 9 \, q_0 ) \, N_3 \over 4 \, q_0^2 } 
+ { 2 \, i \, R \over q_0 \, g^3 }
\right )   
\over
\Gamma \left ( 
1 - { ( 11 + 9 \, q_0 ) \, N_3 \over 4 \, q_0^2 } 
+ { 2 \, i \, R \over q_0 \, g^3 }
\right ) 
\, }
\nonumber \\
&&
\;\;\;\;\;\;\;\;\;\;
\times \;
_1 F_1 \left (
{ ( 11 + 9 \, q_0 ) N_3 \over 4 \, q_0^2 } 
- { 2 \, i \, R \over q_0 \, g^3 }
, \;
{ ( 11 + 9 \, q_0 ) N_3 \over 2 \, q_0^2 } 
, \;
{ 8 \, i \, V \over q_0 \, g }
\right ) \; .
\label{eq:ode_sol1}
\eea
We remind the reader that, by virtue of Eq.~(\ref{eq:q0}), 
in all the above expressions $q_0$ is just a numerical constant, 
$ q_0 \, \equiv \, 2 \, \pi / \cos^{-1} ( \third ) = 5.1043 $.
Note that for weak coupling the curvature terms become
more important due to the $1/g^3 $ coefficient.
The resulting probability distribution
$ \vert \psi (V,R) \vert^2 $ is shown, for some
illustrative cases, in Figures 3,4 and 5.

One important proviso should be be stated here first.
We recall that having obtained an (exact or approximate) 
expression for the wave function does not lead immediately
to a complete solution of the problem.
This should be evident, for example, from the general expression
for the average of a generic quantum operator ${\cal O} (g) $
\beq
\langle {\cal O} (g) \rangle \; \equiv \; 
{ \langle \Psi \vert {\cal O} \vert \Psi \rangle 
\over \langle \Psi \vert \Psi \rangle }
\; = \; 
{ \int  d \mu [g] \cdot {\cal O} ( g_{ij} ) \cdot
\vert \, \Psi [ g_{ij} ] \, \vert^2 
\over 
\int  d \mu [g] \cdot \vert \, \Psi [ g_{ij} ] \, \vert^2 } \; ,
\label{eq:ave_o}
\eeq
where $ d \mu [g] $ is the appropriate (DeWitt) 
functional measure over the three-metric $g_{ij}$.
Because of the general coordinate invariance of the
state functional, the inner products shown above
clearly contain an infinite redundancy due to the geometrical
indistinguishability of 3-metrics which differ
only by a coordinate transformation \cite{dew67}.
Nevertheless this divergence is of no essence
here, since it cancels out between the numerator
and the denominator.

On the lattice the above average translates into
\beq
\langle {\cal O} (l^2) \rangle \; \equiv \; 
{ \langle \Psi \vert {\cal O} \vert \Psi \rangle 
\over \langle \Psi \vert \Psi \rangle }
\; = \; 
{ \int  d \mu [ l^2 ] \cdot {\cal O} ( l^2 ) \cdot
\vert \, \Psi [ l^2 ] \, \vert^2 
\over 
\int  d \mu [ l^2 ] \cdot \vert \, \Psi [ l^2 ] \, \vert^2 } \; ,
\label{eq:ave_o_latt}
\eeq
where $ d \mu [l^2] $ is the Regge-Wheeler lattice 
transcription of the DeWitt functional measure \cite{dew67} 
in terms of edge length variables, here denoted collectively by $l^2$.
The latter includes an integration over all squared
edge lengths, constrained by the triangle inequalities
and their higher dimensional analogs \cite{hw93}.
Again, because of the continuous local diffeomorphism
invariance of the lattice theory,
the individual inner products shown above
will contain an infinte redundancy due to the geometrical
indistinguishability of 3-metrics which differ
only by a lattice coordinate transformation.
And, again, this divergence will be of no essence
here, as it is expected to cancel between numerator
and denominator \cite{har85}.

It seems clear then that, in general, the full functional 
measure cannot be decomposed into a simple product of integrations
over $V$ and $R$.
It follows that the averages listed above are in general still 
highly non-trivial to evaluate.
In fact, quantum averages can be written again quite
generally in terms of an effective (Euclidean) three-dimensional action
\beq
\langle \Psi \vert {\cal \tilde O} (g) \vert \Psi \rangle \; = \; 
{\cal N} \int  d \mu [g] \; {\cal \tilde O} ( g_{ij} ) \;
\exp \left \{ - S_{eff} [g] \right \}  \; ,
\label{eq:ave_seff}
\eeq
with $ S_{eff} [g] \equiv - \ln  \vert \Psi [ g_{ij} ] \vert^2 $ and
${\cal N}$ a normalization constant.
The operator $ {\cal \tilde O} (g) $ itself can be local, or nonlocal as
in the case of correlations such as the gravitational Wilson loop \cite{hw07}.
Note that the statistical weights have zeros corresponding to the nodes of the
wave function $\Psi$, so that $S_{eff}$ is infinite there.
\footnote{
In practical terms, the averages in
Eqs.~(\ref{eq:ave_o}) and (\ref{eq:ave_o_latt}) are difficult
to evaluate analytically, even once the complete wave function is 
known explicitly, due to the non-trivial nature of the
gravitational functional measure;
in the most general case these averages will have to be evaluated
numerically.
The presence of infinitely many zeros in the statistical weights 
complicates this issue considerably, again from a numerical point of view.}

Nevertheless it will make sense here to consider a
{\it semi-classical} expansion for the $3+1$-dimensional theory,
where one simply focuses on the clearly identifiable stationary points 
(maxima) of the probability distribution $\vert \psi \vert $, obtained
by squaring the solution in 
Eqs.~(\ref{eq:ode_sol}) or (\ref{eq:ode_sol1}).
In the following we will therefore focus entirely
on the properties of the probability distribution
$ \vert \psi (V,R) \vert^2 $ obtained from
Eq.~(\ref{eq:ode_sol}) or (\ref{eq:ode_sol1}).
For illustrative purpose, the reader is referred to Figures
3,4 and 5 below.

As discussed previously, the asymptotic expansion 
for the wave function at large volumes is suggestive of
a phase transition at some $G=G_c$
[see for example Eqs.~(\ref{eq:alpha}) and (\ref{eq:G_c})].
In addition, the explicit solution in Eq.~(\ref{eq:ode_sol1}) allows
a more precise non-perturbative characterization of the two
phases.
In view of the non-trivial and generally complex arguments 
of both the gamma function and the confluent hypergeometric
function, the analytic properties of the wave function,
and therefore of the probability distribution,  
are quite rich in features, at least for the more general 
and physically relevant case of non-zero curvature.

One first notes that for strong enough coupling $g$ the 
distribution in curvature is fairly flat around 
$R=0$, giving rise to large fluctuation in the latter
(see Figure 3).
On the other hand, for weak enough coupling $g$ the 
probability distribution in curvature is such that values
around $R=0$ are almost excluded, since they are associated
with a very small probability.
Furthermore, unless the volume $V$ is very small,
the probability distribution is also generally
markedly larger towards positive curvatures (see Figure 4).

In order to explore specifically the curvature ($R$) dependence
of the probability distribution,
it would be desirable to factor out or remove the dependence
of the wave function $\psi (V,R)$ on the total volume $V$.
To achieve this, one can employ a mean-field-type prescription,
and replace the total volume $V$ by its average $ \langle V \rangle $.
After all, the probability distribution in the volume
is well behaved at large $G$ [see Sec.~(\ref{sec:avevol})],
and does not exhibit any
marked change in behavior for intermediate $G$ [as
can be inferred, for example, from the asymptotic
form of the wave function in Eq.~(\ref{eq:psi_largev})].
Consequently we will now make the replacement in $\psi (V,R)$
\beq
V \;\; \longrightarrow \;\; 
\langle \; V \; \rangle \; 
\; \equiv \; N_3 \, \langle \; V_{\sigma} \; \rangle \; = \; 0.2643 \;
{ G \over \sqrt{\lambda} } \; = \; 0.3738 \; g \; ,
\eeq
obtained by inserting the result of Eq.~(\ref{eq:vol_tet}).
This replacement then makes it possible to
plot the wave function of Eq.~(\ref{eq:ode_sol1}) squared
as a function of the coupling $g$ and the total curvature 
$R$ {\it only} (in the following we use again $N_3 = 10$ 
for illustrative purposes); see Figure 5.
One then notes that for strong enough gravitational
coupling $g = \sqrt{G}$ the
probability distribution is again fairly flat around $R=0$,
giving rise to large fluctuations in the curvature.
On the other hand, for weak enough coupling $g$ one
observes that curvatures close to zero have near vanishing
probability.
The distributions shown suggest therefore
what seems a pathological ground state for weak
enough coupling $g < g_c$ [or $G < G_c$, see Eq.~(\ref{eq:G_c})],
with no sensible four-dimensional continuum limit.

At this point some preliminary conclusions, based on the behavior of the 
wave function discussed previously in Sec.~(\ref{sec:curv}) and
the shape shown in Figures 3,4 and 5, are as follows.
For large enough $G > G_c$, but nevertheless close
to the critical point, the flatness in the curvature
probability distribution implies that different 
curvature scales are all equally important.
The corresponding gravitational correlation length is finite
in this region as long as $G>G_c$, and expected to diverge 
at the critical point, thus presumably signaling the presence 
of a massless excitation at $G_c$
[see the argument after Eq.~(\ref{eq:curv_fluc})].
On the other hand for weak enough coupling, $G<G_c$ we
observe that the probability distribution appears to 
change dramatically.
The main evidence for this is the shape of the approximate 
wave function given in Eq.~(\ref{eq:ode_sol}), which
points to a vanishing relative probability for metric field
configurations for which the curvature is small $R \approx 0$.
This would suggest that the weak coupling phase,
for which $G<G_c$, has {\it no} continuum limit in terms of
manifolds that appear smooth, at least at large scales.
The geometric character of the manifold is then inevitably
dominated by non-universal short-distance lattice artifacts;
no sensible scaling limit exists in this phase.

If this is indeed the case, then the results obtained
in the present, Lorentzian, $3+1$ theory generally agree with
what is found in the Euclidean case, where the
weak coupling phase was found to be pathological
as well \cite{hw84,lesh84} (it bears more resemblance 
to a branched polymer, and has thus no sensible 
interpretation in terms of smooth four-dimensional manifolds).
In either case, the only physically acceptable
phase, leading to smooth manifolds at large distances,
seems to be the one with $G > G_c $.
It is a simple consequence of renormalization group
arguments that in this phase the gravitational
coupling at large distances can only flow
to larger values, implying therefore gravitational 
anti-screening as the only physically possible outcome.
Nevertheless it needs to be emphasized here again that
these conclusions have been obtained from a determination
of the wave functional at small curvatures;
it should be clear that when the curvature cannot be regarded as
small, higher order terms in the curvature expansion
of Eqs.~(\ref{eq:psi_asym}) and (\ref{eq:ode_sol}) need to be
retained, which leads us beyond the scope of the present work.

\newpage

\begin{figure}
\begin{center}
%\epsfxsize=7cm
%\epsfbox{.eps}
\includegraphics[width=0.7\textwidth]{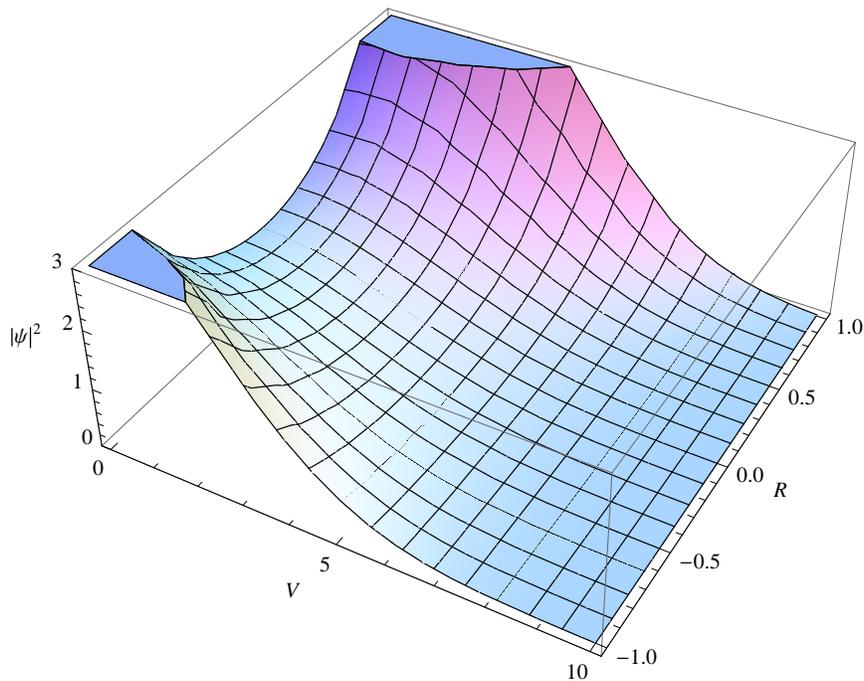}
\end{center}
\caption{\label{fig:psi_s}
Wave function of Eq.~(\ref{eq:ode_sol1}) squared, 
$ \vert \psi (V,R) \vert^2 $, plotted as a function
of the total volume $V$ and the total curvature $R$,
for coupling $g = \sqrt{G} = 1$ and $N_3 = 10$.
One notes that for strong enough coupling $g$ the 
distribution in curvatures is fairly flat around 
$R=0$, giving rise to large fluctuations in the curvature.
These become more pronounced as one approaches the
critical point at $g_c$.
}
\end{figure}

\newpage

\begin{figure}
\begin{center}
%\epsfxsize=7cm
%\epsfbox{.eps}
\includegraphics[width=0.7\textwidth]{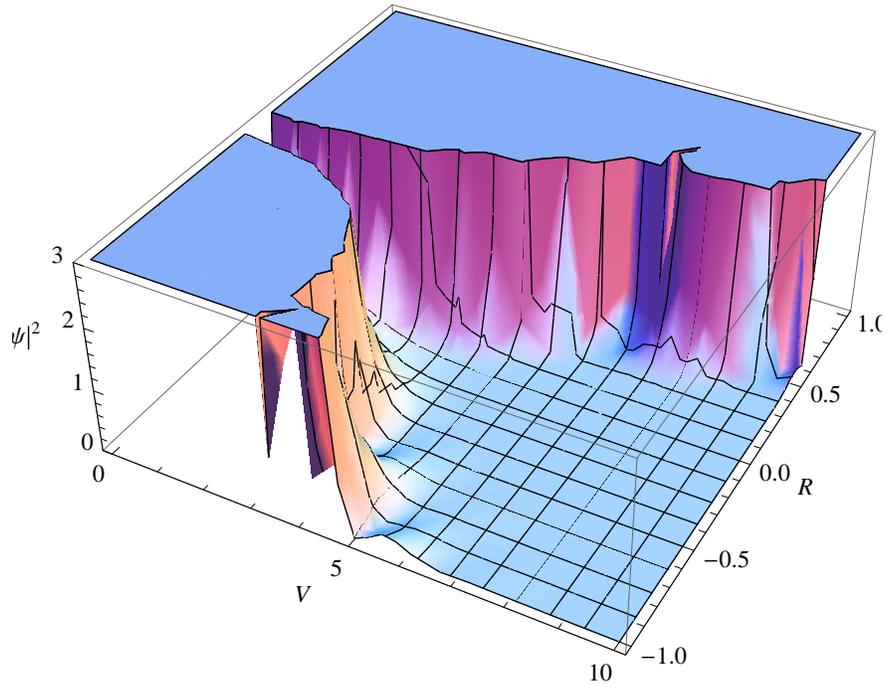}
\end{center}
\caption{\label{fig:psi_w}
Same wave function of Eq.~(\ref{eq:ode_sol1}) squared, 
$ \vert \psi (V,R) \vert^2 $, plotted as a function
of the total volume $V$ and the total curvature $R$,
but now for weaker coupling $g = \sqrt{G} = 0.5$, and still $N_3 = 10$.
For weak enough coupling $g$ the 
distribution in curvature is such that values
around $R=0$ are almost completely excluded, as these 
are associated with a very small probability.
Note that, unless the total volume $V$ is very small,
the probability distribution is markedly larger towards positive curvatures.
}
\end{figure}

\newpage

\begin{figure}
\begin{center}
%\epsfxsize=7cm
%\epsfbox{.eps}
\includegraphics[width=0.7\textwidth]{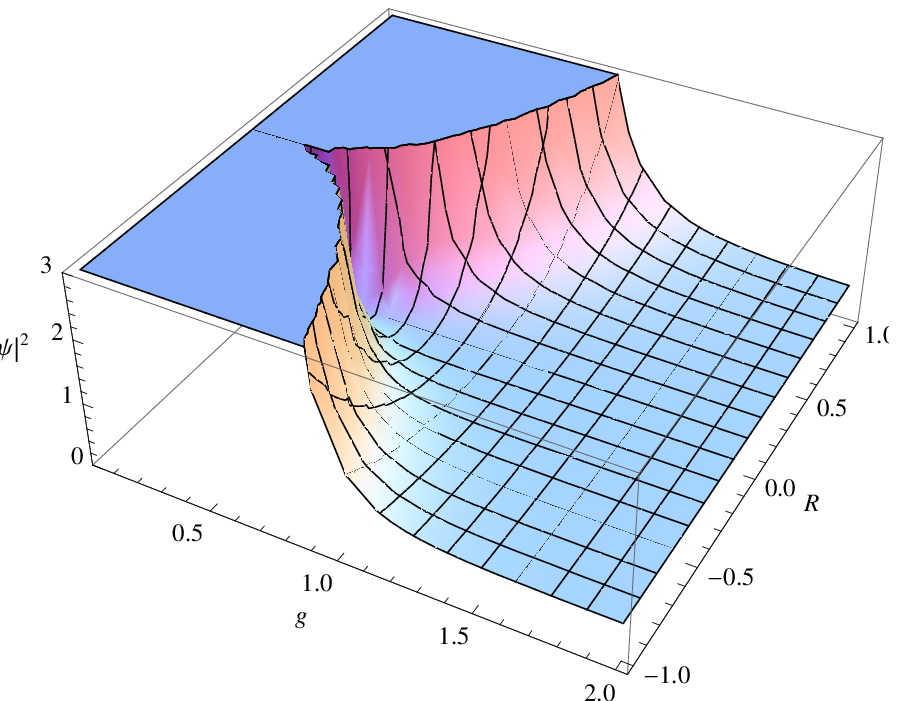}
\end{center}
\caption{\label{fig:psi_g}
Curvature distribution in $R$ as a function of the coupling $g=\sqrt{G}$.
The strong coupling relationship between
the average volume and the coupling $g$
[Eq.~(\ref{eq:vol_tet})] allows one to
plot the wave function of Eq.~(\ref{eq:ode_sol1}) squared
as a function of the coupling $g$ and the total curvature $R$ only
(we use again here $N_3 = 10$ for illustrative purposes).
Then, for strong enough coupling $g = \sqrt{G}$, the
probability distribution $ | \psi |^2 $ is again fairly flat around $R=0$,
giving rise to large fluctuations in the curvature.
The latter are interpreted here as signaling 
the presence of a massless particle.
On the other hand, for weak enough coupling $g$ one
notices that curvatures close to $R=0$ have essentially vanishing
probability.
The distribution shown here points therefore toward
a pathological ground state for weak
enough coupling $g < g_c$ [given in Eq.~(\ref{eq:G_c})],
with no sensible continuum limit.
}
\end{figure}

\newpage

\vskip 40pt

\section{Summary and Conclusions}

\label{sec:sum}

In this work we have discussed the nature of gravitational
wave functions that were found to be solutions of the lattice 
Wheeler-DeWitt equation for finite simplicial lattices.
The main results here were given in Eqs.~(\ref{eq:ode_r}),
(\ref{eq:ode_sol}) and (\ref{eq:ode_sol1}).
While there are many aspects of this problem that still remain
open and unexplored, we have nevertheless shown that the
very structure of the wave function is such that it allows one
to draw a number of useful and perhaps physically relevant 
conclusions about ground state properties of
pure quantum gravity in $3+1$ dimensions.
These include the observation that the theory
exhibits a phase transition at some critical
value of Newton's constant $G_c$ [given in Eq.~(\ref{eq:G_c})].

The structure of the wave function further suggests
that the weak coupling phase, for which the coupling $G<G_c$,
is pathological and cannot be interpreted in terms 
of smooth manifolds at {\it any} distance scale. 
In view of these results it is therefore not entirely surprising
that calculations that rely on the weak field, semiclassical
or small $G$ expansion run into serious trouble and uncontrollable 
divergences very early on.
Such an expansion does not seem to exist if the non-perturbative lattice
results presented here are taken seriously.
The correct physical vacuum apparently cannot in any way be obtained
as a small perturbation of flat, or near-flat, spacetime.
On the other hand the strong coupling phase does
{\it not} exhibit any such pathology, and is therefore a
good candidate for a physically acceptable ground state
for pure quantum gravity.
It is then a simple consequence of standard renormalization
group arguments that in this phase Newton's constant
grows with distance, so that this phase is expected to
exhibit gravitational anti-screening.
Still, to make the problem tractable, most of the results
presented in this work have been obtained in the limit
of small curvatures.
This is clearly a limitation of the present approach.
A more general treatment of the problem, where a variety of curvature 
squared terms are retained in the expansion of the wave functional,
should be feasible by the methods presented here, but
is for now clearly beyond the scope of the present work. 

Let us mention here that in the Euclidean lattice theory of gravity in four dimensions
it was also found early on \cite{hw84,lesh84} [see \cite{ham00} for
more recent numerical investigations of 4d lattice gravity, including
the determination of the critical point and scaling exponents] that
the weak coupling (or gravitational screening) phase is
pathological with no sensible continuum limit, corresponding 
to a degenerate lower dimensional branched polymer.
The calculations presented here
can be regarded, therefore, as consistent with the conclusions reached
earlier from the Euclidean lattice framework.
No new surprises have arisen so far when considering 
the Lorentzian $3+1$ theory, using what can be regarded as
an entirely different set of tools.

It seems also worthwhile at this point to compare with other attempts
at determining the phase structure of quantum gravity
in four dimensions.
Besides the Regge lattice approach, there have been other
attempts at searching for a non-trivial RG UV fixed point 
in four dimensions using continuum methods.
In one popular field theoretic approach one develops a perturbative
diagrammatic $2+\epsilon$ continuum expansion using the background
field method to two loop order \cite{wei79,eps,aid97}.
This then leads to a non-trivial UV fixed point 
$G_c = {\cal O} (\epsilon) $ close to two dimensions.
Two phases emerge, one implying again gravitational screening,
and the other anti-screening.
In the truncated renormalization group calculations for gravity
in four dimensions \cite{reu98,lit04}, where one 
retains the cosmological and Einstein-Hilbert
terms, and possibly later some higher derivative terms,
one also finds evidence for a non-trivial UV fixed point scenario.
As in the case of gauge theories, both of these methods are ultimately 
based on renormalization group flows and the weak field expansion,
and are therefore unable to characterize the non-perturbative 
features of either one of the two ground states.
Indeed, within the framework of the weak field expansion inherent
in these methods, only the weak coupling phase has a chance to start with.
It is nevertheless encouraging that such widely different
methods tend to point in the same direction, namely a 
non-trivial phase structure for gravity in four dimensions.

Let us add here a few more comments, aimed at placing the present work
in a wider context.
Over the years a number of attempts have been made to obtain results
for the gravitational wave functional $\Psi [g]$ in the absence of sources.
Often these have relied on the weak field expansion in
the continuum, see for example \cite{kuc76,kuc92}.
In $3+1$ dimensions one then finds
\beq
\Psi [ h^{TT} ] \; = \; 
{\cal N} \; \exp \left \{ - \quarter
\int d^3 {\bf k} \; k \,
h^{TT}_{ik} ( {\bf k } ) \; 
h^{TT *}_{ik} ( {\bf k } ) \right \} \; ,
\label{eq:psi_weak}
\eeq
where $h^{TT}_{ik} ( {\bf k } ) $ is the Fourier amplitude of
transverse-traceless modes for the linearized gravitational field.
It is clear that the above wave functional describes a collection
of harmonic oscillator contributions, one for each of the 
physical modes of the linearized gravitational field.
It is not necessary to use Fourier modes,
and, as in the case of the electromagnetic field, one can write
equivalently the ground state
wave functional in terms of first derivatives of the field 
potentials,
\beq
\Psi [ h^{TT} ] \; = \; 
{\cal N} \; \exp \left \{ - { 1 \over 8 \pi^2 } 
\int d^3 {\bf x} \int d^3 {\bf y} \; 
{ h^{TT}_{ik,l} ( {\bf x } ) \;
h^{TT *}_{ik,l} ( {\bf y } ) 
\over \vert {\bf x} \, - \, {\bf y} \vert^2 }
\right \} \; .
\label{eq:psi_weak1}
\eeq
Nevertheless, it is generally understood that the above expressions represent only
the leading term in an expansion involving infinitely
many terms in the metric fluctuation $h_{ij}$
(in an expansion about flat space, the cosmological
constant contribution does not appear).
Since Eq.~(\ref{eq:psi_weak}) is just the leading term in the 
weak field expansion,
no issue of perturbative renormalizability appears to this order.
Nevertheless, higher orders are expected to bring in
ultraviolet divergences which cannot be reabsorbed into a simple
redefinition of the fundamental couplings $G$ and $\lambda$.
Then the results presented in this paper [namely Eqs.~(\ref{eq:ode_r}),
(\ref{eq:ode_sol}) and (\ref{eq:ode_sol1})]
can be viewed therefore as a first attempt in extending non-perturbatively
the result of Eq.~(\ref{eq:psi_weak}), beyond the inherent
limitations of the weak field limit.

% NEW text added :
We see a number of additional avenues by which the present work could
be extended.  
One issue, which could be rather laborious to work out in $3+1$ dimensions, 
is the systematic determination of the relevant lattice wave
functionals for the regular triangulations of the sphere 
to higher order in the weak 
field expansion, as was done for example in $2+1$ dimensions \cite{htw12}. 
It should also be possible to obtain the lattice wave functional 
numerically in cases where the triangulation itself is not regular, 
but is described instead by an average coordination number $<\! q \!>$,
as described earlier in the text. Another interesting problem would 
be the derivation of the general form of the lattice wave functional by 
methods which differ from the Frobenius power series method presented
here, such as the $WKB$ approximation \cite{dew67} or the
Raleigh-Schr\"odinger approach. 
Finally, it would also be of some interest to re-derive the form of 
the lattice wave functional for other discrete triangulations, such
as the case of the three-torus $T^3$ (the Kasner model) \cite{htw13}.

%\newpage

\vspace{20pt}

{\bf Acknowledgements}

The work of H.W.H. was supported in part by the Max 
Planck Gesellschaft zur F\" orderung der Wissenschaften, and
by the University of California.
He wishes to thank prof. Hermann Nicolai and the
Max Planck Institut f\" ur Gravitationsphysik (Albert-Einstein-Institut)
in Potsdam for warm hospitality. 
The work of R.M.W. was supported in part by the UK Science 
and Technology Facilities Council. 
The work of R.T. was supported in part by a DED GAANN Student Fellowship.

% \newpage

\vfill

%\newpage

\end{document}